\documentclass{aa}  
\usepackage{graphicx}
\usepackage{txfonts}
\usepackage{multirow}
\RequirePackage{tikz}
\begin{document}

   \title{Radiative transfer in cylindrical threads with incident radiation
   \thanks{Appendix figures A1-A5 are only available in electronic form via
http://www.edpsciences.org}
}

   \subtitle{VII. Multi-thread models}

   \author{N. Labrosse
          \and
          A.S. Rodger
                    }
   \authorrunning{Labrosse \& Rodger}

   \institute{SUPA, School of Physics and Astronomy, University of Glasgow, Glasgow G12 8QQ, Scotland\\
              \email{Nicolas.Labrosse@glasgow.ac.uk}
             }

   \date{}

  \abstract
   {}
   {Our aim is  to {improve on previous radiative transfer calculations in illuminated cylindrical threads  to better understand} the physical conditions in cool solar chromospheric and coronal structures {commonly observed in hydrogen and helium lines}.}
   {We solved  the radiative transfer and statistical equilibrium equations in a two-dimensional cross-section of a cylindrical structure {oriented horizontally and lying above the solar surface}. The cylinder is filled with a mixture of hydrogen and helium and is illuminated at a given altitude from the solar disc.  {We constructed simple models made from a single thread or from {an ensemble} of several threads along the line of sight. This first use of two-dimensional, multi-thread fine structure modelling  combining hydrogen and helium radiative transfer allowed us to compute} synthetic emergent spectra from {cylindrical} structures {and to study the effect of line-of-sight integration of an ensemble of threads} under a range of physical conditions. We analysed the effects of variations in  temperature distribution and {in} gas pressure. We considered the effect of multi-thread structures within a given field of view and the effect of peculiar velocities between the structures in a multi-thread model. We compared these new models  to the single {thread} model and tested them with varying parameters.}
   {{The presence of a temperature gradient, with temperature increasing towards the edge of the cylindrical thread, reduces the relative importance of the incident radiation coming from the solar disc on the emergent intensities of most hydrogen and helium lines.}
   We {also} find that when assuming randomly displaced {threads} in a given field of view, the integrated intensities  of optically thick and thin transitions behave considerably differently. In optically thin lines, the emergent intensity increases proportionally with the number of {threads}, and the spatial {variation of the} intensity  becomes increasingly homogeneous. Optically thick lines, however,   saturate after only a few {threads}. As a consequence, the spatial {variation of the} intensity retains much similarity with that of the first few threads. The multi-thread model {produces} complex line profiles with significant asymmetries if randomly generated line-of-sight velocities are added for each {thread}. }
   {{These new computations show, for the first time, the effect of integrating the radiation emitted in H and He lines by several cylindrical threads that are static or moving along the line of sight. They can be used to interpret high-spatial and spectral resolutions of cylindrical structures found in the solar atmosphere, such as cool coronal loops or prominence threads.}}


   \maketitle
%

\section{Introduction}
\label{sec:intro}

To model cool {horizontal} structures in the solar atmosphere (e.g. cool coronal loops, fine {filament, or} prominence threads), we use a two-dimensional cylindrical radiative transfer model departing from local thermodynamic equilibrium (NLTE). A two-dimensional cylinder here refers to a cylindrical model with only two active variables out of the three cylindrical coordinates. The development of this model is described in a series of papers (\citealt[{hereafter papers I to VI}]{2004A&A...413..733G,2005A&A...434.1165G,2006A&A...448..367G,2007A&A...465.1041G,2008A&A...487..805G,2009A&A...503..663G}). 
Below, we  summarise  the results of the most relevant studies in this series \citep[see also][]{2015ASSL..415..131L}.

\citet[{paper I}]{2004A&A...413..733G} started with a basic one-dimensional model for a pure hydrogen structure (a vertical cylinder in the solar atmosphere). \citeauthor{2004A&A...413..733G} investigated  two accelerated lambda-iteration (ALI) methods with the aim to test their strengths and to create an accurate, efficient numerical model.  
The model {was} then expanded in \citet[{paper II}]{2005A&A...434.1165G} to a more general two-dimensional, azimuth-dependent model. The code {was then} capable of simulating variations in the radiation field from both radial and azimuthal coordinates. The incident radiation on the cylinder {comes} from a uniformly radiating sphere at a given distance and inclination. 

\citet[{paper III}]{2006A&A...448..367G}   {combined} the results from the first {two} papers to create a working model for a 10-level plus continuum hydrogen atom {with all lines calculated using \emph{complete redistribution in frequency} (CRD). An arbitrary inclination of the cylinder with respect to the solar surface can be chosen. The author investigated the formation of several Lyman lines and the Lyman continuum under different temperature and pressure conditions, including two-dimensional (2D) distributions of temperature and pressure to model intensities emitted by complex structures. The behaviours of line profiles and integrated intensities were discussed}. 

{Next, \citet[{paper IV}]{2007A&A...465.1041G} and \citet[{paper V}]{2008A&A...487..805G} added various improvements such as time-dependent solutions to study thermal equilibrium and the effects of 3D velocity fields on the emergent hydrogen spectrum, respectively.}

Finally, \citet[{paper VI}]{2009A&A...503..663G} expanded on the {work} described in paper III {and added a \ion{He}{i}--\ion{He}{ii}--\ion{He}{iii} system to model chromospheric and coronal loop-like structures visible in hydrogen and helium lines. In Paper~VI, {the} electron density is determined by the ionisation equilibria of both hydrogen and helium}. In that paper the authors successfully studied the effects of temperature, pressure, and helium abundance using isothermal and non-isothermal {isobaric} models. The  models described there present a good method for modelling single thread structures. {However, in paper~VI there was no attempt to investigate} the  effects of several threads {present along} the line of sight (LOS) on the emergent spectra.

High spatial resolution observations clearly show that prominences are made of very fine threads \citep[see, for example,][]{2008ApJ...676L..89B, 2006SoPh..234..115C,  2007A&A...472..929G, 2005SoPh..226..239L, 2012ApJ...747..129L,  2014LRSP...11....1P, 2014A&A...569A..85S,  2015ASSL..415.....V}. The {number} of small, fine structure threads in close proximity to each other is likely greater beyond the resolution of current instrumentation.
{Similarly, cool loops in the solar atmosphere are usually observed to be in groups.}
 Hence, it is important to be able to model the radiation emerging from a bundle of threads.

\cite{2007A&A...472..929G} used a 2D {thread model in Cartesian geometry} to compute the emergent radiation in hydrogen lines, and extended the modelling {to a multi-thread configuration} to represent a series of threads along the LOS. The authors investigated the emission in the hydrogen Lyman lines from an arbitrary number of {threads}, each with a given prominence-to-corona transition region. By comparing their computed line profiles to observations from the SUMER instrument on SOHO, they concluded that the spectra produced from prominence fine structures are better simulated with a {multi-thread} model as compared to a single {thread} model. 
Their model assumes  magnetohydrostatic (MHS) equilibrium \citep{2001A&A...375.1082H}. This allows them to infer realistic pressure profiles across the {2D threads}. This in turn determines any density fluctuations along the LOS. 

Line profiles of optically thick transitions often display a substantial central reversal, where the intensity at the central wavelength is significantly absorbed by the material in the LOS. The resulting profile becomes double peaked. Various effects can cause these two peaks to become asymmetric, for example either the red or blue peak becomes higher in intensity with respect to the other. These asymmetric line profiles have been observed frequently in solar prominences and their fine structure for the hydrogen Lyman lines with the SUMER instrument aboard SOHO by \cite{2001A&A...370..281H,1999SoPh..189..109S}, and \cite{2007SoPh..241...53S}, and earlier with the OSO-8 (\emph{Orbiting solar Observatory}) mission by \cite{1982ApJ...253..330V}.

In \cite{2008A&A...490..307G}, the authors expand on their previous model to include random LOS velocities for each of the {threads}. They found that the LOS velocities produce asymmetries in the H Lyman line profiles that agree with SUMER {prominence} observations. 
{In Paper~V, a single thread was considered with varying plasma velocity distributions (radial, tangential, and longitudinal flows). The different types of velocity distributions were found to have different effects on the synthetic hydrogen line  profiles. Plasma oscillations in single thread with Cartesian geometry have been successfully modelled in \cite{2014A&A...562A.103H}. This study finds that different oscillatory modes produce spectral indicator variations with differing magnitudes and that a seismology  analysis of the hydrogen H$\alpha$ and H$\beta$ line parameters could be used to find and diagnose oscillatory modes in solar prominences.} 
{It should be noted that the LOS velocities used in this study are global quantities and only vary between threads in multiple thread models.} 

{Several review articles have been written on solar prominence observations and modelling. \cite{2007ASPC..368..271H}  gives a review of fine structure of solar prominences, and solar cool loop observations are discussed in \cite{2014LRSP...11....1P}}. \cite{2010SSRv..151..243L} provides an overview of the concepts and techniques used in NLTE radiative transfer modelling applied to prominences (e.g., a solar structure illuminated from the solar disc).

Our aim is  to {present new calculations that can} improve the understanding of the physical conditions in cool solar chromospheric and coronal structures.  
To accomplish this, we detail  the models we use and the results we obtain in the following sections. Section~\ref{sec:models} starts by providing the main features of the modelling. Section~\ref{sec:multiloop_results} presents the set-up for considering the superposition of several static individual threads and the results from that investigation. Section~\ref{sec:vlos} details how velocities are introduced in the modelling. Finally, Section~\ref{sec:conclusions} offers some conclusions about this work.
 
\section{Modelling}\label{sec:models}

In this section we briefly summarise the {main} features of the models {that we have used}. The modelling details are given in {Paper~VI}.
The model atom for hydrogen  contains five discrete energy levels and a
continuum. This allows for ten discrete transitions and five bound-free transitions. The helium
model atom, on the other hand, has 34 discrete states; with 29 for \ion{He}{i}, four for \ion{He}{ii}, and one for {\ion{He}{iii}}. The number of permitted transitions {for the entire helium system} is 76.  Under usual conditions,
a large proportion of these are optically thin. The
models for both the hydrogen and helium atoms assume CRD for all discrete transitions. 

{\subsection{Parameters defined for each single thread}}

The inclination of a cylindrical thread is defined by the angle between the axis of the cylinder and the vertical, $\alpha$. Owing to symmetry, $\alpha$ is considered between 0 and $90\degr$. We consider {cylindrical threads} oriented horizontally in the solar atmosphere, i.e. $\alpha = 90\degr$. The LOS intersects the cylinder perpendicularly to the cylinder axes. {For a full schematic view of the model geometry , see Figs.~1 and 2 in Paper~II.}

The  size  of the {thread} {(i.e. the diameter of the cross-section of the cylindrical thread) can be chosen to represent}  a large-scale prominence modelled as a whole (thickness $10^3$--$10^4$~km), or a significantly smaller fine structure of thickness $\sim10^2$~km. {We focus on threads of diameter $1000-2000$~km.}

The important physical parameters within the thread are gas pressure $P_\mathrm{g}$, temperature $T$, helium abundance, $A_\mathrm{He}$, and microturbulent velocity. The effects of abundance and of microturbulent velocity are  not investigated here. In all models the microturbulent velocity is 5~km $\mathrm{s^{-1}}$ {with a helium abundance ratio of $A_{\mathrm{He}}$=0.1}.

To understand the effects of gas pressure and temperature on the overall system, two groups of {models are used}: \emph{t} {models (isothermal and isobaric)} and \emph{p} models {(isobaric models with a radial temperature distribution)}. {The same models} were first used in Paper~III, {as well as} in Paper~VI, {and we therefore use the same notation here}. 

\begin{figure*}
\centering
                \includegraphics[width=\textwidth]{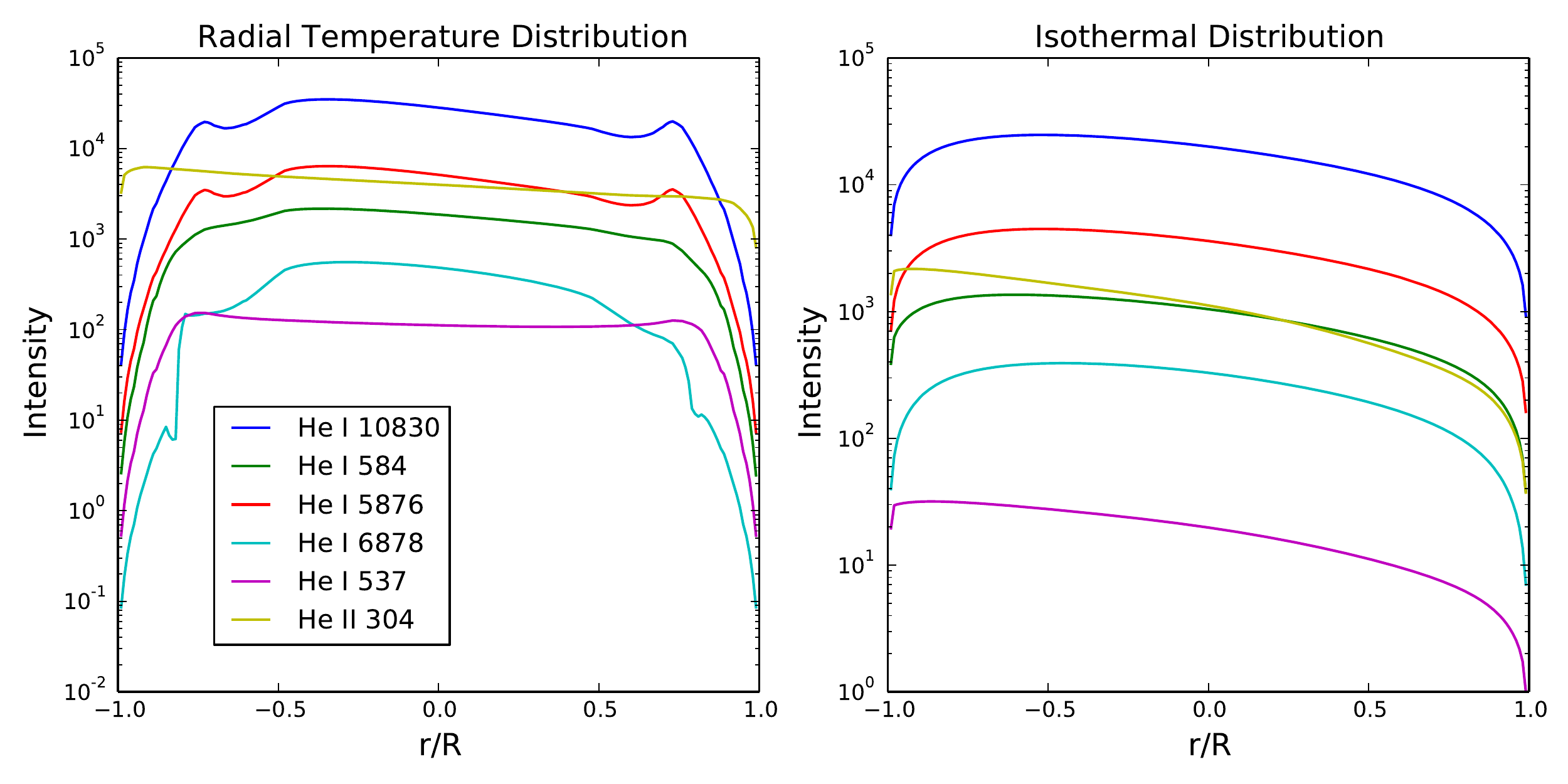}
                \caption{Effect of incident radiation on helium line intensity across a section of a {cylindrical thread}. The frequency-integrated intensity is measured in erg $\mathrm{cm^{-2}}$ $\mathrm{s^{-1}}$ $\mathrm{sr^{-1}}$. The position is {given in fractional units of the cylinder radius}. The {horizontal} axis {represents the position} vertically away from the solar surface with the zero point at the {thread} centre. {The left-hand graph used the \emph{p}4 model, whilst the right-hand graph used the isothermal \emph{t}1 model.}}
                \label{fig:incident}
\end{figure*}

{All of the \emph{t} models in this paper  use a constant gas pressure of 0.1 dyn cm$^{-2}$}. By varying the isothermal temperature in these models, the effect of large-scale temperature variations can be {investigated}. 
It is, however, often necessary to take the {presence of a distribution of temperatures into account}, which is expected to {result in} a significant temperature gradient, especially near the cylinder edges. {The importance of the transition region in the case of prominences was discussed by \cite{1999A&A...349..974A}.} The profile of the temperature variation is not {well} known, {though it is possible to make some assumptions. \cite{1999A&A...349..974A} proposed an analytical expression of the temperature and pressure depending on the column mass inside a 1D quiescent prominence model. This was extended to 2D by \cite{2001A&A...375.1082H}. Here,} the temperature distribution chosen to model this change {in \emph{p} models} is {the same} as in Paper~III and is {given by}
\begin{equation}\label{eq:tempdist}
        \log{T(r)} = \log{T_0} + (\log{T_1} -\log{T_0}) \frac{r-r_0}{r_1-r_0} \ .
\end{equation}
The parameter $T_0$ defines the temperature at the centre of the thread and $T_1$ represents the temperature of the surrounding corona. The inner and outer edges of the {transition region} are defined by $r_0$ and $r_1$, respectively. {The  temperature at the thread centre is $6\times 10^3$~K, whilst the  temperature at the coronal boundary is $10^{5}$~K.} This distribution is not created from a theoretical basis and is designed only to depict the effect of a radial temperature gradient. {A plot showing such a temperature variation across the thread can be found in Fig.~1 of Paper~VI.} 

The cylinder diameter is 1000~km in \emph{t} models and 2000~km in \emph{p} models {(e.g. for our \emph{p} models, $r_0=500$~km and $r_1=1000$~km). This choice makes  the effect of adding a temperature gradient on top of the isothermal thread core clearer when comparing the results between \emph{t} and \emph{p} models}.
The  altitude of the cylinder influences how the radiation coming from the solar disc is illuminating the structure (through, most notably, the height-dependent dilution factor). We use a fixed height of 10000~km above the solar surface. 
The  effect that incident radiation may have on the  {emergent intensities depends on the plasma parameters and on the lines under consideration. For instance, cool isothermal threads are comparatively more sensitive to the incident radiation than non-isothermal models. Indeed, the incident radiation determines the boundary conditions for the solution of the radiative transfer equation.} 

{For \emph{p} models, the excitation and ionisation state of the plasma may be primarily driven by the gas  temperature, which can be large at the outer edge of the thread illuminated by the incident radiation}. In a model with a {typical} temperature distribution, the effect of incident radiation from the Sun or other sources can easily be masked by the local temperature gradients. This effect can be seen in Fig.~\ref{fig:incident}, which compares the frequency-integrated emergent intensity in several helium lines for an isothermal model and a model with a temperature gradient. {Fig.~\ref{fig:incident} shows that in an isothermal model (\emph{t}1, right panel), the variation of the emergent integrated intensity along a vertical cut through the cross-section of the cylinder is primarily driven by the change in the radiation incident on the structure as we go from the bottom to the top of the horizontal thread. On the other hand, the variation of the emergent integrated intensities in most lines for a model with a temperature gradient (\emph{p}4, left panel) shows a lesser difference between the bottom and top edges of the cylindrical thread with local variations due to plasma conditions. However, the \ion{He}{ii} 304~\AA\ line still shows the influence of the anisotropic incident radiation on the cylindrical thread, since this line is mostly formed by the scattering of the incident radiation under the physical conditions considered here.}

The {model} parameters we used  are summarised in Table~\ref{table:parameters}. {We frequently use the \emph{p}4 model since it corresponds to a typical gas pressure (0.1 dyn~cm$^{-2}$), which is also adopted in all \emph{t} models, making comparisons between these models easier.} 

\begin{table}
\caption{Parameters for \emph{p} {(with temperature gradient)} and \emph{t} {(isothermal)} models: {gas pressure, temperature, and cylinder diameter}.}             
\label{table:parameters}      
\centering                          
\begin{tabular}{c c c c}        
\hline\hline                 
Model & $P_\mathrm{g}$ (dyn cm$^{-2}$) & $T$ ($10^3$ K) & $D$ (km)\\     
\hline                        
\multirow{2}{*}{\emph{p}1 -- \emph{p}7}  & \{0.02, 0.03, 0.05,  & $T$(centre)=6 & \multirow{2}{*}{2000}\\
                        & 0.1, 0.2, 0.3, 0.5\} & $T$(surface)=100\\[1em] 

\multirow{2}{*}{\emph{t}1 -- \emph{t}11}  & \multirow{2}{*}{0.1} & \{6, 8, 10, 15, 20, 30 & \multirow{2}{*}{1000}\\
                        & & 40, 50, 65, 80, 100\} \\
\hline                                   
\end{tabular}
\end{table}

\subsection{Multi-thread models}\label{sec:multiloop_method}

This article aims to {apply the methods} of the multi-thread  model of \cite{2007A&A...472..929G} {in Cartesian geometry} to the cylindrical {models} of Paper~VI. In doing this, the effect of multi-thread modelling is also applied to helium lines, which has been relatively unexplored so far {\citep[for a preliminary study see][]{2009A&A...498..869L}}. Our focus is on investigating the vertical cross-sectional intensity, to try and discern any important geometrical effects. The method used to do this is described {below}, and the results are given in Section~\ref{sec:multiloop_results}.

The multi-thread model considers an arbitrary number of cool threads, $N$, oriented horizontally in the solar atmosphere with an inclination angle $\alpha$ equal to $90\degr$. In the single thread model, the field of view was automatically set by the physical boundaries of the cylinder in consideration, i.e. the edge closest and furthest from the Sun in the vertical direction. In the multi-thread case {this definition is kept, using the closest thread to the observer} (hereafter referred to as the foremost thread). 

In the multi-thread model, the final emergent intensity after $N$ threads is calculated as the sum of the constituent intensities from each individual thread. However, photons emitted by the furthest thread ($N^{\mathrm{th}}$) have to travel through all other threads along the LOS ($N - 1$ threads). Its intensity is hence reduced by the cumulative effect of all the other threads optical thicknesses. The equation for the total emergent intensity is similar to the following equation used by \cite{2007A&A...472..929G}:
\begin{equation}\label{eq:multi_loop} 
        I_{total} = I_1 + I_2 \mathrm{e}^{-\tau_{1}} + \dots + I_N \mathrm{e}^{-\sum^{\mathrm{N}-1}_{1} \tau_i} \ ,
\end{equation}
where $I_{total}$ is the frequency and position-dependent total emergent intensity from $N$ threads; and $I_{i}$ and $\tau_{i}$ are the frequency and position-dependent emergent intensity and optical thickness of the $i^{\mathrm{th}}$ thread, respectively. The number 1 here is used to represent the foremost thread.  

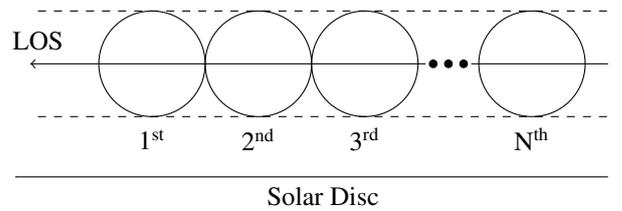
\begin{figure}
        \begin{center}
                \begin{tikzpicture}
                        \draw [dashed](0,0.7) -- (7.5,0.7);
                        \draw [<-] (-0.1,0) -- (5.1,0);
                        \draw [dashed](0,-0.7) -- (7.5,-0.7);
                        \draw (1.5,0) circle (0.7);
                        \draw (2.9,0)circle (0.7);
                        \draw (4.3,0)circle (0.7);
                        \draw [fill=black] (5.2,0) circle (0.05);
                        \draw [fill=black] (5.4,0) circle (0.05);
                        \draw [fill=black] (5.6,0) circle (0.05);
                        \draw [-] (5.7,0) -- (7.5,0);
                        \draw (0,0.3) node {LOS};
                        \draw (6.5,0) circle (0.7);
                        \draw (1.5,-1) node {$1^{\mathrm{st}}$};
                        \draw (2.9,-1) node {$2^{\mathrm{nd}}$};
                        \draw (4.3,-1) node {$3^{\mathrm{rd}}$};
                        \draw (6.5,-1) node {$\mathrm{N}^{\mathrm{th}}$};
                        \draw [-] (-0.3,-1.5) -- (7.5,-1.5);
                        \draw (3.75,-1.75) node {Solar Disc};
                \end{tikzpicture}
        \end{center}
        \caption{Schematic diagram showing layout of simple fully aligned multi-thread model. The dashed lines represent the upper and lower edges of the field of view, with the lower edge facing the solar disc.}
        \label{fig:simple_case}
\end{figure}

After a brief inspection of Eq.~(\ref{eq:multi_loop}), it is is easy to see that there is a special case of this scenario, in which all threads are identical and perfectly aligned with respect to each other and the LOS (Fig.~\ref{fig:simple_case}). In this case, for each frequency and position across the cross-section of the {cylinder}, all $I_{i}$ and $\tau_{i}$ are  identical. In this case, the total emergent intensity is greatly simplified as
\begin{equation}\label{eq:simple_case}
        I_{total} = I \sum^{i = \mathrm{N}-1}_{i = 0} \mathrm{e}^{-i \tau} \ ,
\end{equation}
where $I$ and $\tau$ are the identical intensities and optical thicknesses from each single thread in the simplified multi-thread model. 

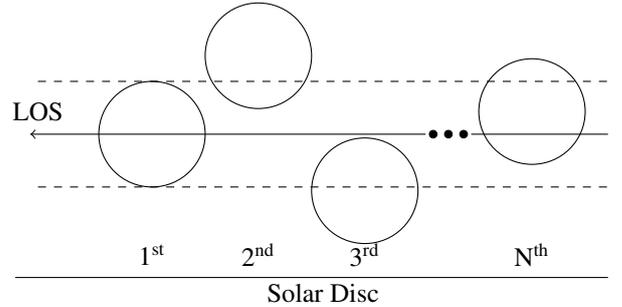
\begin{figure}
        \begin{center}
                \begin{tikzpicture}
                        \draw [dashed](0,0.7) -- (7.5,0.7);
                        \draw [<-] (-0.1,0) -- (5.1,0);
                        \draw [dashed](0,-0.7) -- (7.5,-0.7);
                        \draw (1.5,0) circle (0.7);
                        \draw (2.9,1.04)circle (0.7);
                        \draw (4.3,-0.75)circle (0.7);
                        \draw [fill=black] (5.2,0) circle (0.05);
                        \draw [fill=black] (5.4,0) circle (0.05);
                        \draw [fill=black] (5.6,0) circle (0.05);
                        \draw [-] (5.7,0) -- (7.5,0);
                        \draw (0,0.3) node {LOS};
                        \draw (6.5,0.3) circle (0.7);
                        \draw (1.5,-1.6) node {$1^{\mathrm{st}}$};
                        \draw (2.9,-1.6) node {$2^{\mathrm{nd}}$};
                        \draw (4.3,-1.6) node {$3^{\mathrm{rd}}$};
                        \draw (6.5,-1.6) node {$\mathrm{N}^{\mathrm{th}}$};
                        \draw [-] (-0.3,-1.9) -- (7.5,-1.9);
                        \draw (3.75,-2.10) node {Solar Disc};
                \end{tikzpicture}
        \end{center}
        \caption{Schematic diagram showing layout of a randomly {positioned} (unaligned) multi-thread model. The dashed lines represent the upper and lower edges of the field of view, {defined by the first thread along the LOS}, with the lower edge facing the solar disc.}
        \label{fig:unaligned}
\end{figure}

The simple fully aligned model is useful as it gives the ability to verify the validity of the multi-thread expansion through an easy comparison with the single thread model. It is however unrealistic, so the less trivial case where the threads are randomly unaligned with respect to each other has to be considered. In this case, the field of view was arbitrarily chosen to again encompass the upper and lower bounds of the foremost thread. Each thread after the foremost thread was given a randomly generated vertical displacement with respect to the foremost thread, which effectively has a displacement of 0. The total width of the field of view was taken to correspond to a distance of one diameter of the cylindrical thread, 1000~km for \emph{t} models, and 2000~km for \emph{p} models. To ensure that at least some of each {thread} was within the field of view, the randomly generated displacement value was capped at +3/4 and -3/4 of {the diameter of the thread}. The values for the displacements {in position array points} used in the model run are \{0, -119, -40, 120, 9, -45, 55, -86, 37, 3\}, {where the total width of the field of view is 201 points across, centred at 0}.  A schematic diagram for this set up can be seen in Fig.~\ref{fig:unaligned}.

{\subsection{Assumptions}}

In creating the simple fully aligned case and the randomly unaligned, multi-thread models there are several important assumptions for the system that should be noted. As mentioned previously, our assumptions for the simple case require that each {thread} in the model has identical physical properties (temperature distribution, gas pressure, etc). There is  no need for this requirement in the unaligned model, although this study is limited to identical threads. 

Another assumption is that, in both cases, the incident radiation on each thread must be identical. That is to say that radiation from the Sun must appear uniform across the distance between the first thread and the last in the LOS. This places limits on both the maximum length of the LOS and/or the minimum altitude of the individual threads. This assumption is particularly important in the unaligned case as it requires the altitude displacement between the threads to be significantly small when compared to the {altitude of the field of view above the solar surface}. 

The final assumption {we make} is that the threads within the multi-thread models are not radiatively interacting. This means that no incident radiation from one thread upon another is considered. This assumption makes the model much simpler as it keeps the equation for the emergent intensity in the form of Eq.~(\ref{eq:multi_loop}) as well as negating any relevance of the horizontal distance between the threads. This assumption is acceptable as the intensity of the incident radiation of the Sun should be considerably larger than any possible incident radiation from one thread to another.\\

\section{Static multi-thread configurations}
\label{sec:multiloop_results}

{\subsection{Fully aligned threads}}

\begin{figure}
                        \includegraphics[width=0.49\hsize]{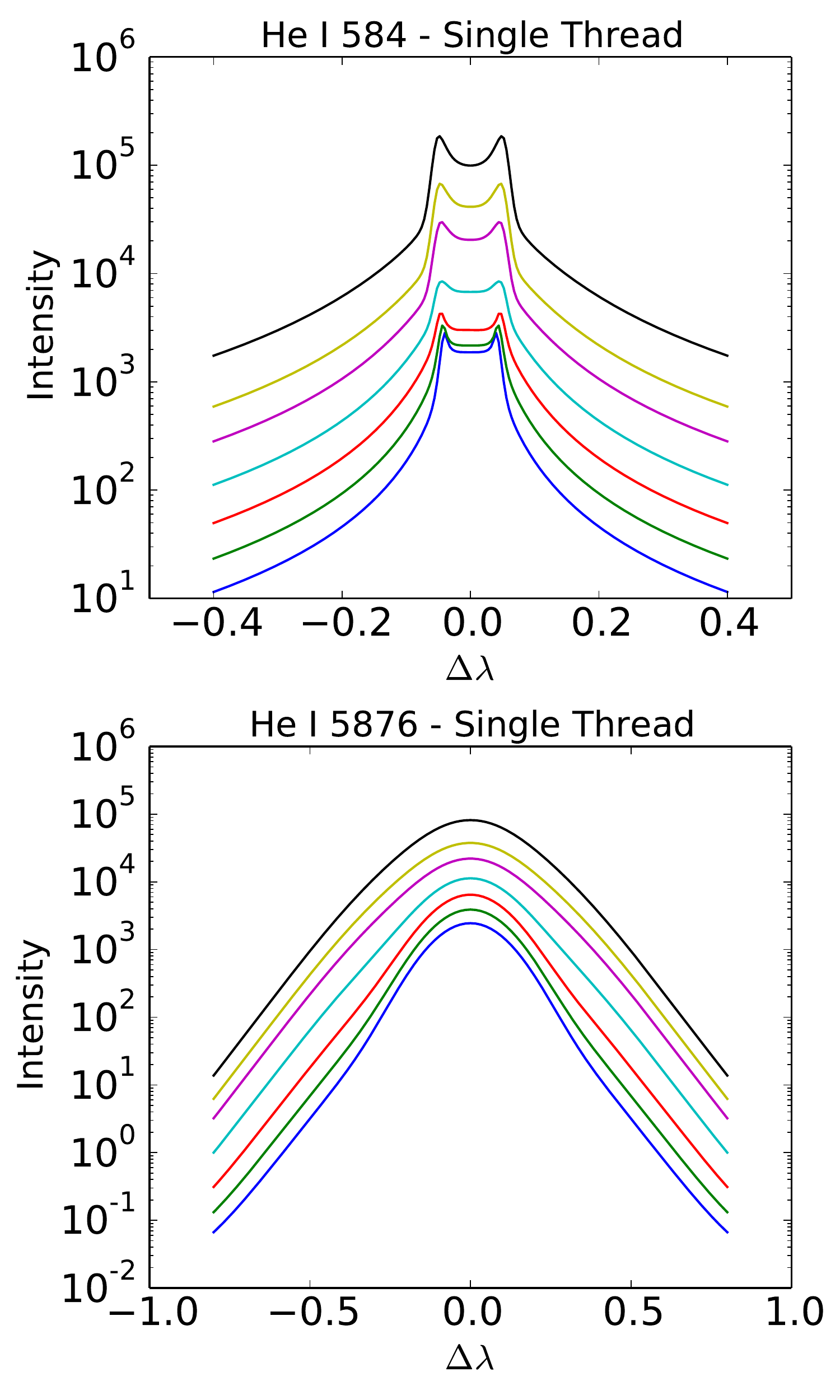}
                        \includegraphics[width=0.49\hsize]{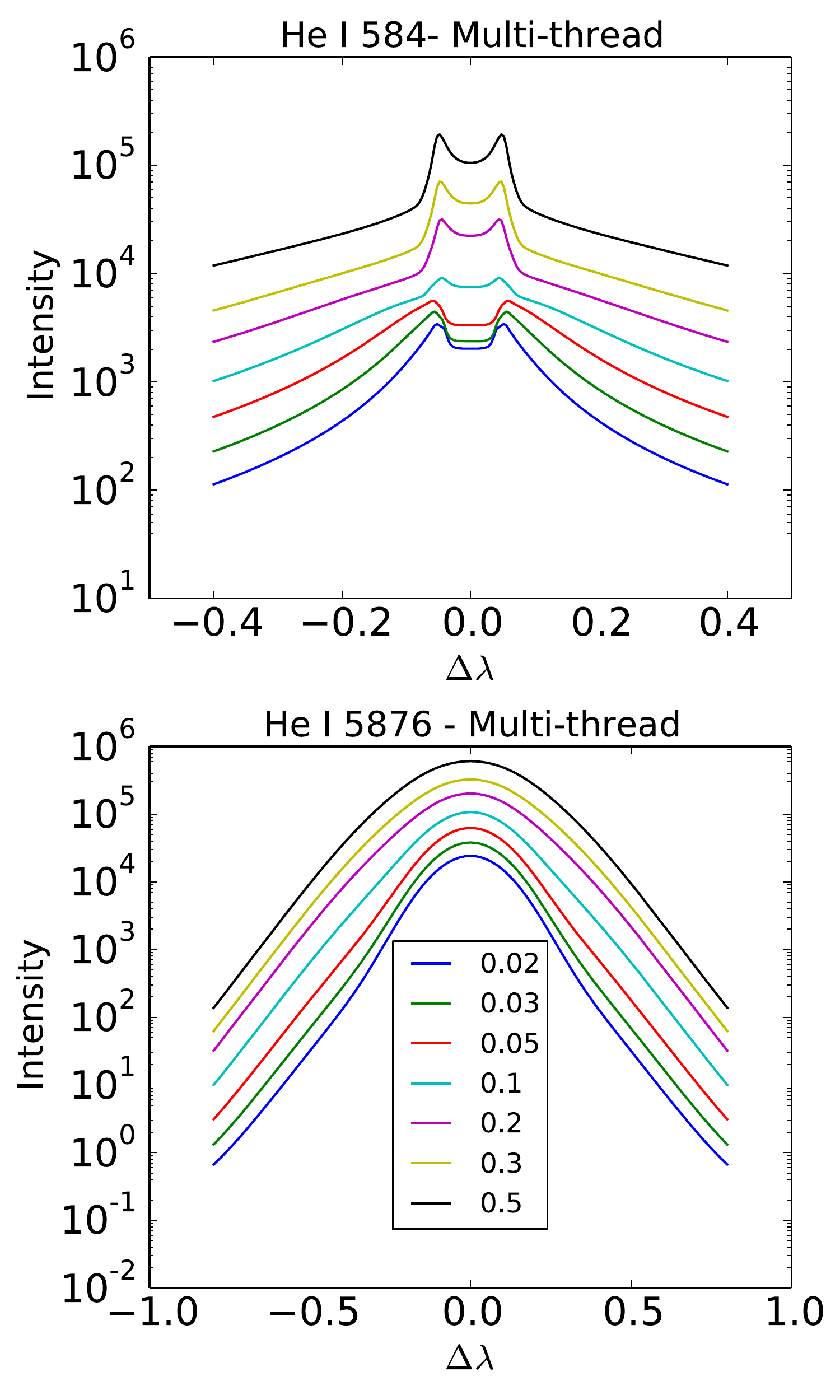}
                \caption{Changes in line profiles for helium lines \ion{He}{i} 584~\AA\ and \ion{He}{i} 5876~\AA\ for a single thread model and for a multi-thread model with ten perfectly aligned threads. Intensities are measured in erg $\mathrm{s^{-1}~cm^{-2}~sr^{-1}~\AA^{-1}}$. The legend gives the isobaric gas pressure in units of dyn $\mathrm{cm^{-2}}$. Delta lambda is measured in \AA.} 
                \label{fig:single_vs_multi}
\end{figure}

The multi-thread {configuration} was first {investigated} by comparing the synthetic spectra  {for ten perfectly aligned threads using the \emph{p}4 model} with the previously {studied} single thread spectra. This was done by taking the position-averaged line spectra from the single thread model and comparing it with the profiles produced by the fully aligned, multi-thread model. The change in \ion{He}{i} 584~\AA\ and \ion{He}{i} 5876~\AA\ line profiles from the single and multi-thread models can be seen in Fig.~\ref{fig:single_vs_multi}. These graphs show the changes in line profiles with gas pressure, where the gas pressure was isobaric across the cylinder as dictated by models p1 through p7 (see Table~\ref{table:parameters}). 

From Fig.~\ref{fig:single_vs_multi} it is clear  that the relationship between line profile shape and gas pressure does not change significantly between a single thread and an ensemble of ten threads for  either  helium line. However, the shape of \ion{He}{i} 5876~\AA\ is also unaffected by the change from a single to a multi-thread model. This is because the plasma is optically thin for \ion{He}{i} 5876~\AA\ photons {(see Table~\ref{table:optical_thicknesses})}: the emergent intensity from $N$, fully aligned threads, is simply  $\sim N$ times the intensity of the single thread. 

The shape of the line profiles for \ion{He}{i} 584~\AA\ changes significantly between the single and multi-thread models. This is because \ion{He}{i} 584~\AA\ is optically thick at line centre, {whilst remaining optically thin in the wings}. Its high optical thickness at line centre prevents photons at that wavelength from travelling far within the medium; they are scattered very efficiently. This causes the line centre to saturate as only the light from the foremost thread emerges. The line however becomes increasingly less optically thick further away from the line centre. This causes the intensity of the {optically thin} line wings to scale proportionally with the number of threads, as seen in \ion{He}{i} 5876~\AA. The combined effects of high optical thickness at line centre and lower optical thickness in the wings creates the profiles seen for \ion{He}{i} 584~\AA. This is similar to what is observed in other optically thick lines. Table 2 shows the optical thicknesses for all transitions we considered. The values of the optical thickness {at line centre and total optical thickness} are averaged across all points in the field of view obtained from a single thread \emph{p}4 model.

\begin{table}
        \caption{Position-averaged {total optical thickness ($\tau$) and optical thickness at line centre ($\tau_0$)} for given transition lines in single thread \emph{p}4 model.}             
\label{table:optical_thicknesses}      
\centering                          
\begin{tabular}{c c c}
\hline\hline
Line & $\tau_0$ & $\tau$\\
\hline
Ly$\alpha$ & 2.15E+05 &1.10E+04\\
Ly$\beta$ & 3.45E+04 & 2.97E+03\\
Ly$\gamma$ & 1.20E+04 & 9.79E+02\\
H$\alpha$ & 4.31E-01 & 1.01E-01\\
H$\beta$ & 5.96E-02 & 1.29E-02\\
P$\alpha$ & 5.93E-03 & 1.99E-03\\
\ion{He}{i} 10830 & 8.30E-02 & 1.94E-02\\
\ion{He}{i} 584 & 1.60E+04 & 5.08E+02 \\
\ion{He}{i} 5876 & 8.12E-03 & 1.54E-03\\
\ion{He}{i} 6878 & 6.61E-04 & 1.23E-04\\
\ion{He}{i} 537 & 3.90E+03 & 2.28E+02\\
\ion{He}{ii} 304 & 1.33E+03 & 6.23E+01 \\
\hline
\end{tabular}
\end{table}

{\subsection{Randomly positioned threads}}

The  multi-thread model was then {studied} for ten randomly positioned threads with  the displacement values given in Section~\ref{sec:multiloop_method}. When compared with the simple, {fully aligned} case {(Fig.~\ref{fig:single_vs_multi})}, the {position-averaged line} profile shapes and intensity values from the unaligned, multi-thread model {are not greatly affected}. This is not unexpected since the displacements were randomly generated and the line profile shape only changes significantly towards the very edge of the {thread}. 

{The clearer effects of the multi-thread configuration with randomly positioned threads were found by plotting the position-dependent intensity across the vertical field of view. For each position in this vertical cross-section, the intensity was integrated over frequency. Figure 5 shows the change in intensity across the field of view,} with increasing numbers of randomly unaligned {(\emph{p}4 model)} threads, for several different hydrogen and helium lines.  

\begin{figure*}[t]
        \begin{center}
                \includegraphics[width=\textwidth]{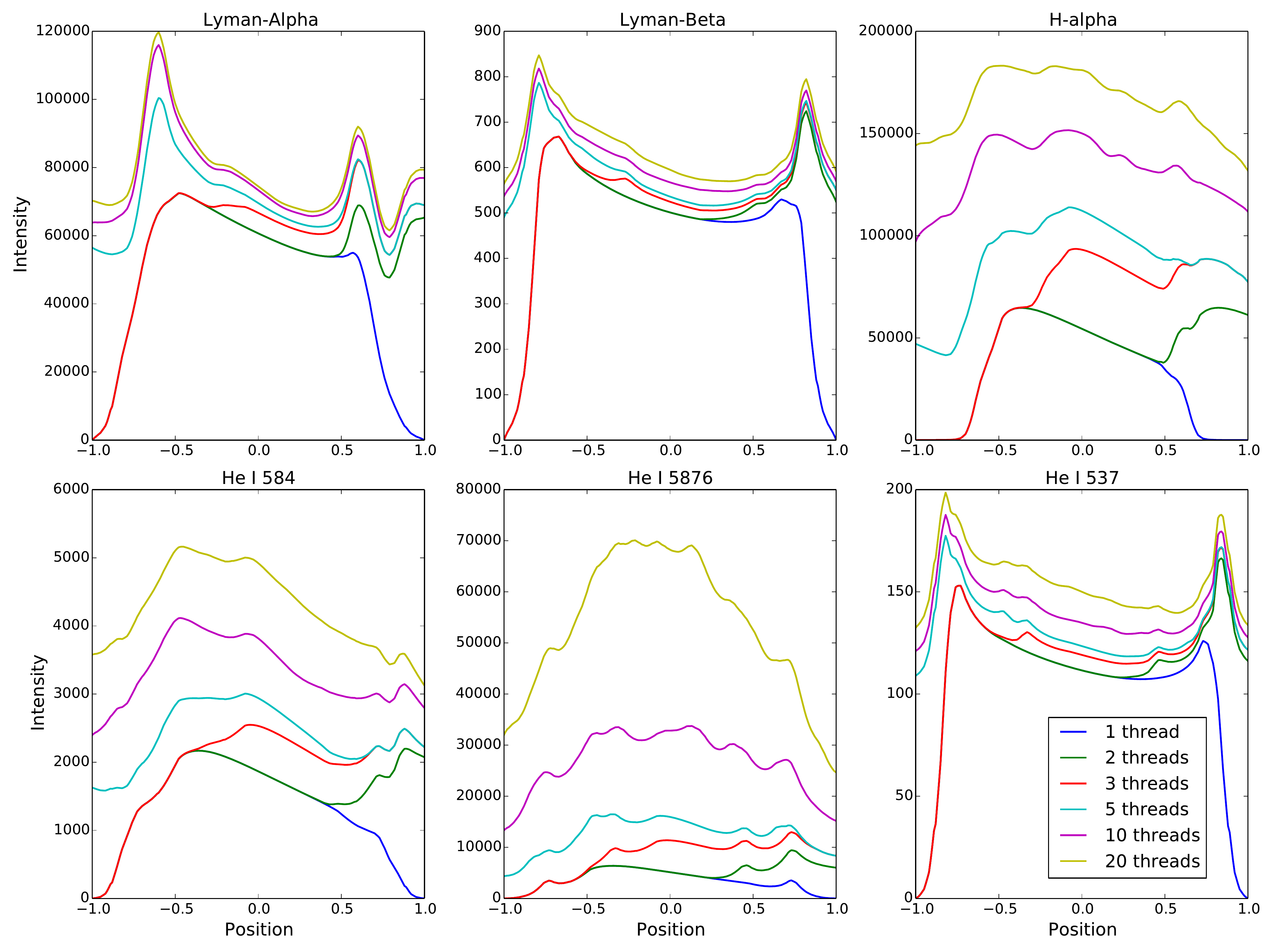}
                \caption{Changes in frequency-integrated intensity of various hydrogen and helium lines {as a function of distance along the vertical direction through the field of view defined by the first thread (from bottom to top), for different}   numbers of randomly unaligned threads (using the \emph{p}4 model). The intensity is given in erg $\mathrm{s^{-1}~cm^{-2}~sr^{-1}}$ and the {vertical} position in the field of view is given in Mm. {The horizontal axis represents the position vertically away from the solar surface with the zero point at the centre of the first thread.}}
                \label{fig:numloops_Xsection}
        \end{center}
\end{figure*}

It is clear from Fig.~\ref{fig:numloops_Xsection} that {the variation of} the frequency-integrated intensity {along the vertical direction} can have a noticeably different behaviour when the number of threads in the field of view is  increased, depending on the particular transition in consideration. Optically thin lines increase in intensity with  increasing numbers of threads, and any {variation of the} intensity along the vertical direction resulting from the presence of several randomly-positioned threads along the LOS {is smoothed} towards homogeneity. {This effect is seen in H$\alpha$, H$\beta$ (not shown in figure), and \ion{He}{i} 5876~\AA.}

In contrast, in optically thick lines only the photons from the closest threads reach the observer, {while photons coming from threads further away along the LOS are lost through multiple scattering}. This saturation effect allows the structure of the nearest threads in the field of view to be maintained,  {independent of the number of  threads that may} lie {further away in the LOS}. This means that comparisons between observations in optically thick lines and optically thin lines could  be used to infer the number of thread structures in the LOS. Figure~\ref{fig:numloops_Xsection} shows the optically thick Lyman $\alpha$, Lyman $\beta$, and \ion{He}{i} 537~\AA\ lines all show {this} saturation {effect} after only a few {threads}. However, \ion{He}{i} 584~\AA, which is an optically thick line, does not display this {behaviour}. This is discussed further in Section~\ref{sec:saturation_pressure}.\\

\subsubsection{{Variation of integrated intensity  along the vertical direction as a function of  pressure}}\label{sec:saturation_pressure}

\begin{figure*}
        \begin{center}
                \includegraphics[width=0.9\textwidth]{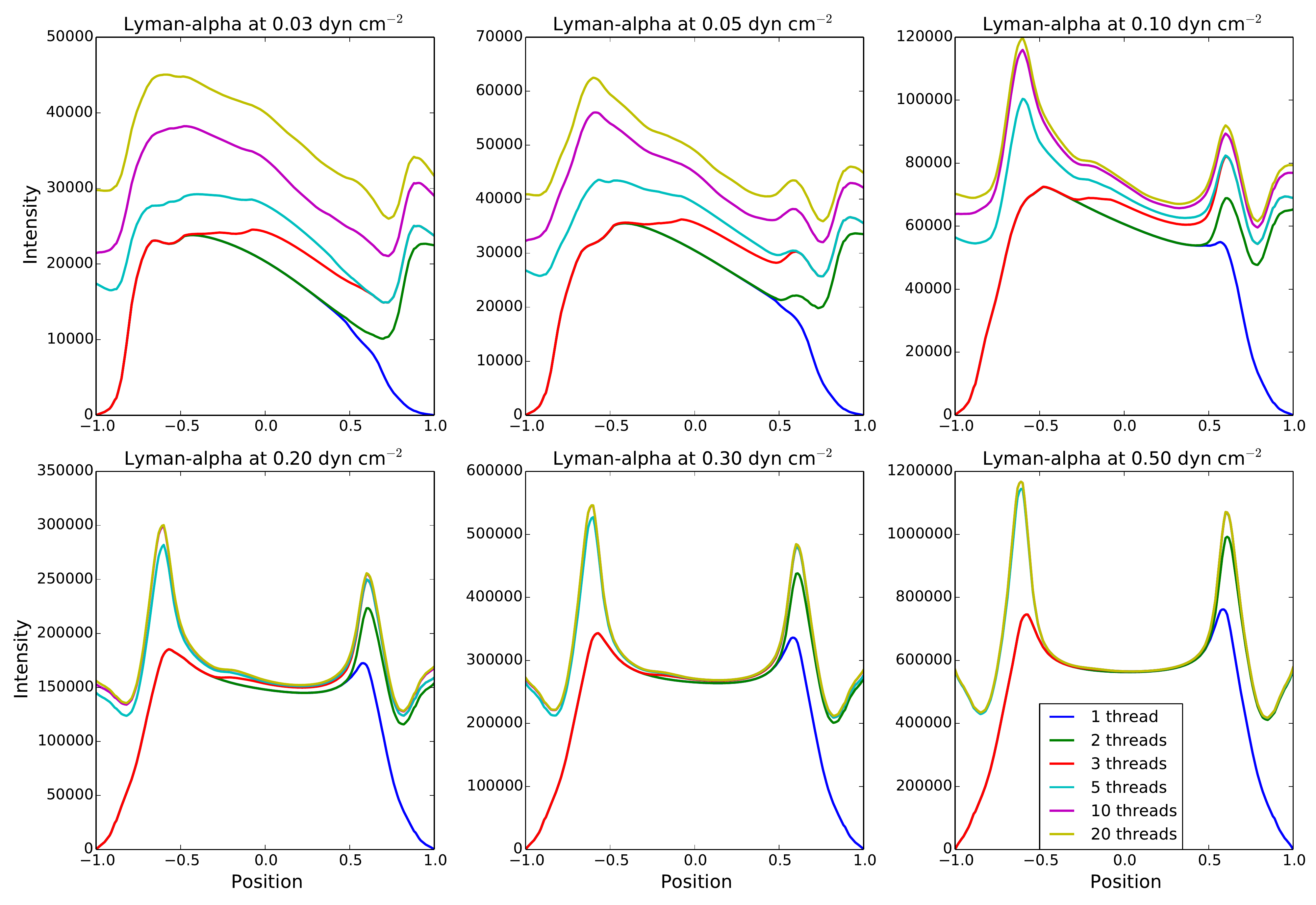}
                \includegraphics[width=0.9\textwidth]{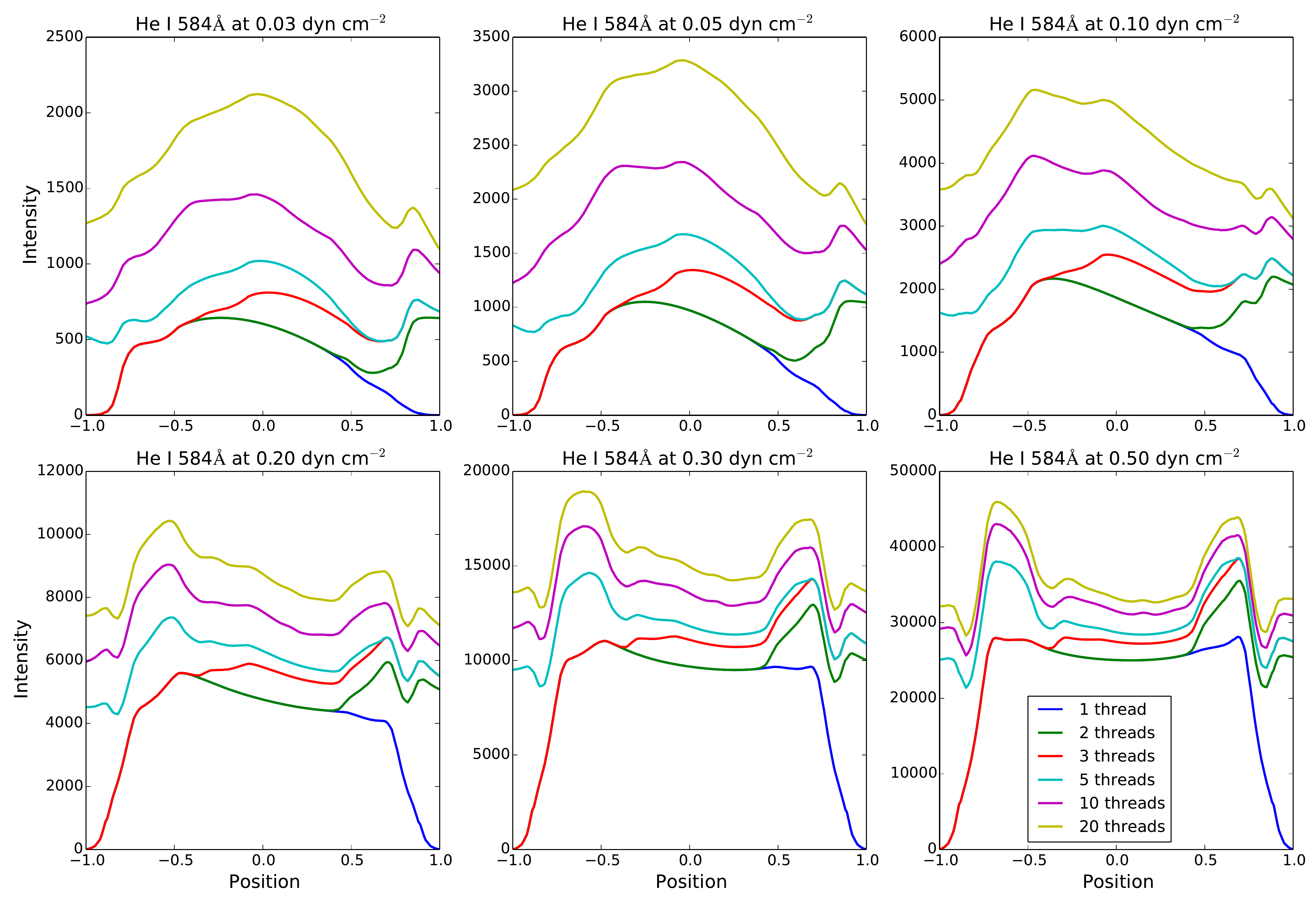}
                \caption{Changes in cross-sectional frequency-integrated intensity of the hydrogen Lyman $\alpha$ line (top two rows) and the \ion{He}{i} 584~\AA\ line (bottom two rows) with an increasing number of unaligned threads for isobaric models \emph{p}1--\emph{p}7. The intensity is given in erg $\mathrm{s^{-1}~cm^{-2}~sr^{-1}}$ and the  position in the field of view is given in Mm. {The horizontal axis represents the position vertically away from the solar surface with the zero point at the centre of the first thread.}}
                \label{fig:saturation_pressures}
        \end{center}
\end{figure*}

Here we look in more details at why some optically thick lines, most prominently \ion{He}{i} 584~\AA, appear to show less integrated intensity saturation with increased number of threads than other optically thick lines. {We suggest} that this is an effect of the line's relationship with gas pressure. In Fig.~\ref{fig:numloops_Xsection} we used the \emph{p}4 model. This model's radial temperature distribution is described {by} Eq.~(\ref{eq:tempdist}) and the gas pressure is fixed at 0.1~dyn cm$^{-2}$. Several optically thick lines begin to show a change in behaviour around that value of the pressure; this can be seen, for example for \ion{He}{i} 584~\AA\ in Fig.~\ref{fig:single_vs_multi}. {A similar change can be seen clearly in Lyman $\alpha$ and to a lesser extent in Lyman $\gamma$ and \ion{He}{i} 537~\AA}. At lower pressures, radiation scattering processes are dominant and increasing the pressure has little effect: the line centre is saturated. At larger pressures, electron collisional processes take a larger role in the formation of the line, and the line centre intensity begins to increase with pressure. 

To {verify} this, the change in frequency-integrated intensity {along the vertical direction} with increasing number of  threads was compared for all the \emph{p}1--\emph{p}7 models. Figure~\ref{fig:saturation_pressures} shows the change in  intensity with pressure for the hydrogen Lyman~$\alpha$ line (top two rows), while the bottom two rows show the results for the \ion{He}{i} 584~\AA\ line.

At low pressures, the integrated intensity does not saturate significantly, and hence much of the foremost thread structure is lost, and {the intensity variation along the vertical direction is smoothed by the addition of more threads}. As pressure increases, the {variation of the} intensity along the vertical direction reaches saturation with fewer and fewer threads. At high pressures the integrated intensity almost immediately saturates, masking the contribution of all of the inner threads. This effect is seen in all optically thick lines, including \ion{He}{i} 584~\AA, however the effect is most pronounced in the hydrogen Lyman $\alpha$ line, {especially at low pressures}.

\begin{figure*}
        \begin{center}
                \includegraphics[width=\textwidth]{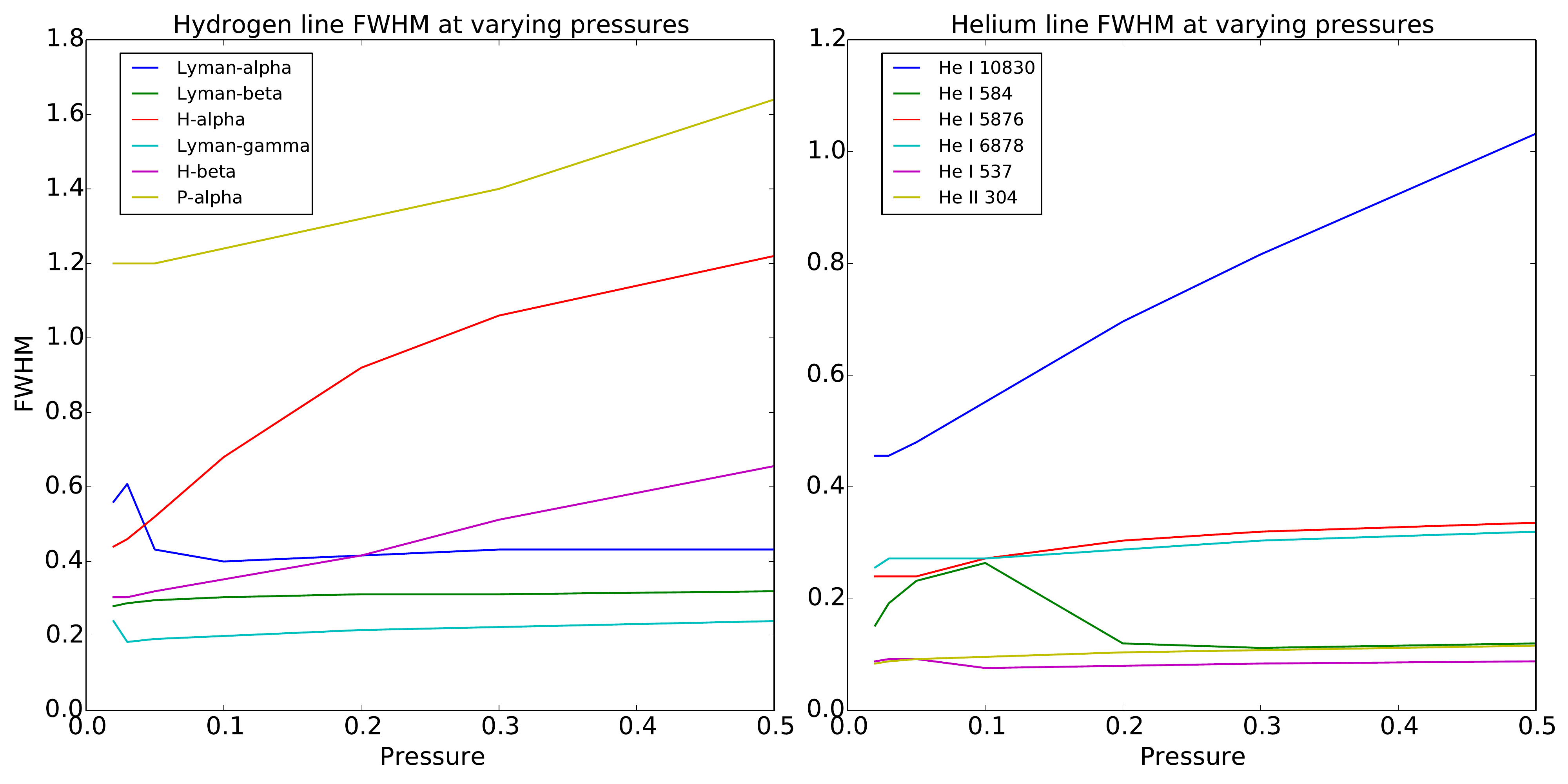}
                \caption{Changes in full width at half maximum (FWHM) with pressure {(models \emph{p}1--\emph{p}7)} for various hydrogen and helium lines. The FWHM is given in \AA\ and the  pressure is given in dyn $\mathrm{cm^{-2}}$.} 
                \label{fig:FWHM}
        \end{center}
\end{figure*}

The change in {behaviour} with pressure is {related to} the comparative widths of the emission line profiles at these pressures. The change in line width with pressure was {investigated} by calculating the full width at half maximum (FWHM). The FWHM was calculated simply as the width at the point where the line presents half its maximum intensity, which may or may not be the central peak (in case the line has a central reversal). The change in FWHM with pressure for {several} hydrogen and helium lines can be seen in Fig.~\ref{fig:FWHM}.

In general, {the intensities of optically thick} lines that have a lower FWHM saturate more  than broader lines. {Broad optically thick lines have extended wings where the optical thickness is much lower than at line centre, and which  then allows the radiation from additional threads along the LOS to be detected. For narrower optically thick lines, only the foremost threads can be detected. As shown in Fig.~\ref{fig:FWHM}, the increased FWHM for Lyman $\alpha$ and \ion{He}{i} 584~\AA\ at low pressures explains their apparent lower saturation, however, at higher pressures their FWHM decreases and their intensity across the field of view saturates.} 

The effect {discussed here} may form a spectral diagnostic tool, as the degree of {variation of the integrated intensity} in the field of view for optically thick lines gives an indication of the number of threads in the system {if the gas pressure is previously known. This is the case as long as  the pressure values for all threads are not too different. The pressure variation across each thread does not have a huge impact on the effect discussed here. From an observational point of view, however, the usefulness of this diagnostic is limited by the spatial resolution of the instrument}.  

\subsubsection{Effect of the number of threads on intensities in  optically thick lines}

\begin{figure}
                \includegraphics[width=\columnwidth]{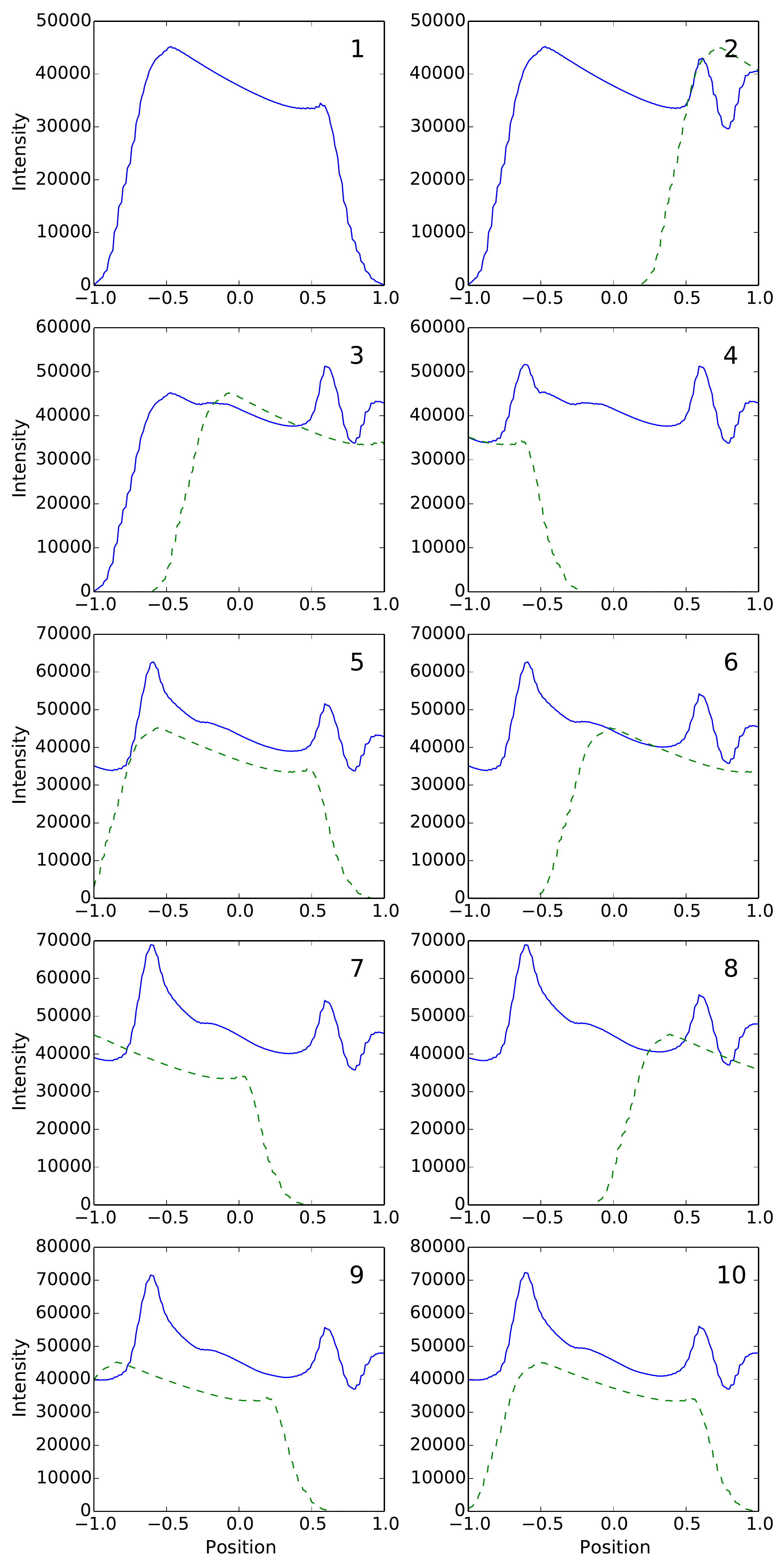}
                \caption{Cumulative effect of unaligned threads on the total emergent intensity  for the Lyman $\alpha$ line. The solid blue line shows the emergent intensity after {integration through} $N$ threads, whilst the dashed green line shows the {emergent intensity} of the $N$-th thread in the LOS{ on its own}. {$N$ is given by the number in the top right-hand corner of each panel.} The intensity is given in erg $\mathrm{s^{-1}~cm^{-2}~sr^{-1}}$ and the position is in Mm. The position direction is oriented vertically away from the solar surface with the zero point at the centre of the field of view. {The radiation emergent from each thread is computed with the \emph{p}4 model.}}
                \label{fig:cumulative_Xsection}
\end{figure}

{As discussed in the previous section,}
 the {variation of emitted intensities of optically thick lines with vertical direction in a multi-thread configuration quickly} saturates, {and the features in the spatial intensity {variation} due to the specific physical conditions in the threads along the LOS are smoothed out}. This allows the structure of the closest threads in the LOS to be reflected in the {spatial variation of the} intensity, regardless of the total number of threads. This structure depends on the locations of the nearest threads in the LOS and most prominently on the foremost thread. The creation of this structure is best seen by comparing the cumulative effect of each thread on the total {field of view after $N$ threads}, an example of this effect can be seen in Fig.~\ref{fig:cumulative_Xsection}. {In Fig.~\ref{fig:cumulative_Xsection} the solid blue lines show the combined effect of $N$ unaligned threads, whilst dashed green lines show the individual emission from the $N^{\mathrm{th}}$ thread. {Each successive thread is added to the LOS behind the foremost thread.} Many of the panels of Fig.~\ref{fig:cumulative_Xsection} illustrate that significant parts of the structural features shown in the blue combined view follow the same patterns as those shown by the foremost thread given in panel 1. Most recognizable features in the variations of the integrated intensity computed using the unaligned multiple thread models  come from the foremost and closest threads in the LOS.}

\section{{Multi-thread configurations with small-scale velocities along the LOS}}\label{sec:vlos}

{As discussed in Section~\ref{sec:intro}, line profiles of optically thick transitions often display a substantial central reversal together with asymmetries in the wings, where one intensity peak is stronger than the other.}
Asymmetric profiles of the hydrogen Lyman lines have been modelled by \cite{2008A&A...490..307G}, who assumed the asymmetries occur because of the presence of random peculiar velocities between individual threads. That is to say that each thread in a multi-thread system is moving with a given velocity with respect to all the other threads. If a proportion of a thread velocity is directed in the LOS, the emergent spectrum from that thread is Doppler shifted by an amount proportional to the magnitude of the velocity in the LOS direction. LOS velocities in prominence fine structures can occur as a result of several different mechanisms, including flows of plasma along the prominence's magnetic field lines or by oscillations of the magnetic structure itself {\citep{2008A&A...490..307G,2008A&A...487..805G,2014A&A...562A.103H}}.   

For LOS velocities to have any effect on the symmetry of the emergent line profiles, a multi-thread system is required. The reason for this is that when considering a thread in isolation, any Doppler shift {along} the LOS simply moves the profile in either the blue or red direction. In a multi-thread system, {photons emitted by a moving thread are ``seen'' by other threads at a different frequency, and therefore at a different optical thickness than if they had been emitted by a thread at rest.} This effect results in the formation of significantly asymmetrical profiles from only fairly small velocities, {as demonstrated by} \cite{2008A&A...490..307G}. The work from these authors  shows that {observed} asymmetries in the line profiles of the hydrogen Lyman lines can be reproduced with a two-dimensional, multi-thread model {in Cartesian geometry}. 

This section of the study aims to describe how the concept of random LOS velocities can be extended to work in a cylindrical {multi-thread} model for both hydrogen and helium line transitions.

\subsection{Adding LOS velocities}\label{sec:vlos_method}

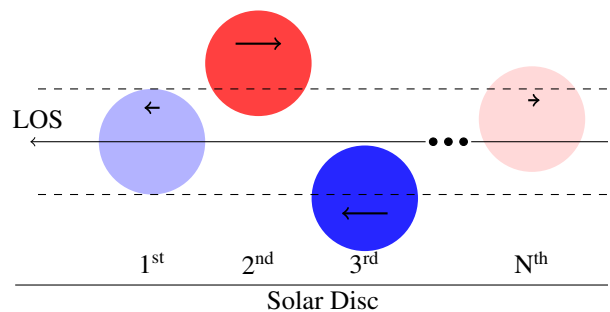
\begin{figure}
        \begin{center}
                \begin{tikzpicture}
                        \fill [blue!30] (1.5,0) circle (0.7);
                        \fill [red!75] (2.9,1.04)circle (0.7);
                        \fill [blue!85] (4.3,-0.75)circle (0.7);
                        \draw [fill=black] (5.2,0) circle (0.05);
                        \draw [fill=black] (5.4,0) circle (0.05);
                        \draw [fill=black] (5.6,0) circle (0.05);
                        \draw (0,0.3) node {LOS};
                        \fill [red!15] (6.5,0.3) circle (0.7);
                        \draw (1.5,-1.6) node {$1^{\mathrm{st}}$};
                        \draw (2.9,-1.6) node {$2^{\mathrm{nd}}$};
                        \draw (4.3,-1.6) node {$3^{\mathrm{rd}}$};
                        \draw (6.5,-1.6) node {$\mathrm{N}^{\mathrm{th}}$};
                        \draw [-] (-0.3,-1.9) -- (7.5,-1.9);
                        \draw (3.75,-2.10) node {Solar Disc};
                        \draw [-] (5.7,0) -- (7.5,0);
                        \draw [thick,<-] (1.4,0.45) -- (1.6,0.45);
                        \draw [thick,->] (2.6,1.3) -- (3.2,1.3);
                        \draw [thick,<-] (4,-0.95) -- (4.6,-0.95);
                        \draw [thick,->] (6.45,0.55) -- (6.6,0.55);                     
                        \draw [dashed](0,0.7) -- (7.5,0.7);
                        \draw [<-] (-0.1,0) -- (5.1,0);
                        \draw [dashed](0,-0.7) -- (7.5,-0.7);

                 \end{tikzpicture}
                 
        \end{center}
        \caption{Layout of unaligned multiple threads with random LOS velocities. The dashed lines represent the upper and lower edges of the field of view with the lower edge facing the solar disc. The arrows within the threads correspond to the LOS velocities, whilst the colour of the thread represents its Doppler shift, i.e. red shift or blue shift.}
        \label{fig:vlos_diagram}
\end{figure}

The  cylindrical multi-thread model, described in Section~\ref{sec:multiloop_results}, was expanded to include randomised peculiar velocities along the LOS between each thread. The velocities described in this section are specifically global velocities, which are experienced identically across the entirety of each thread and are not local flows within the structure. 
The velocity given for each thread is constant across the whole cylinder. The shift in wavelength caused by the resulting Doppler shift from the LOS velocities is $\Delta \lambda_i = \lambda_0 u_i/c$, where $\Delta \lambda_i$ is the Doppler shift in wavelength, $\lambda_0$ is the central wavelength for a given transition, $u_i$ is the velocity component in the LOS direction for the $i^{\mathrm{th}}$ thread, and $c$ is the speed of light.

\begin{figure}
        \begin{center}
                \includegraphics[width=\hsize]{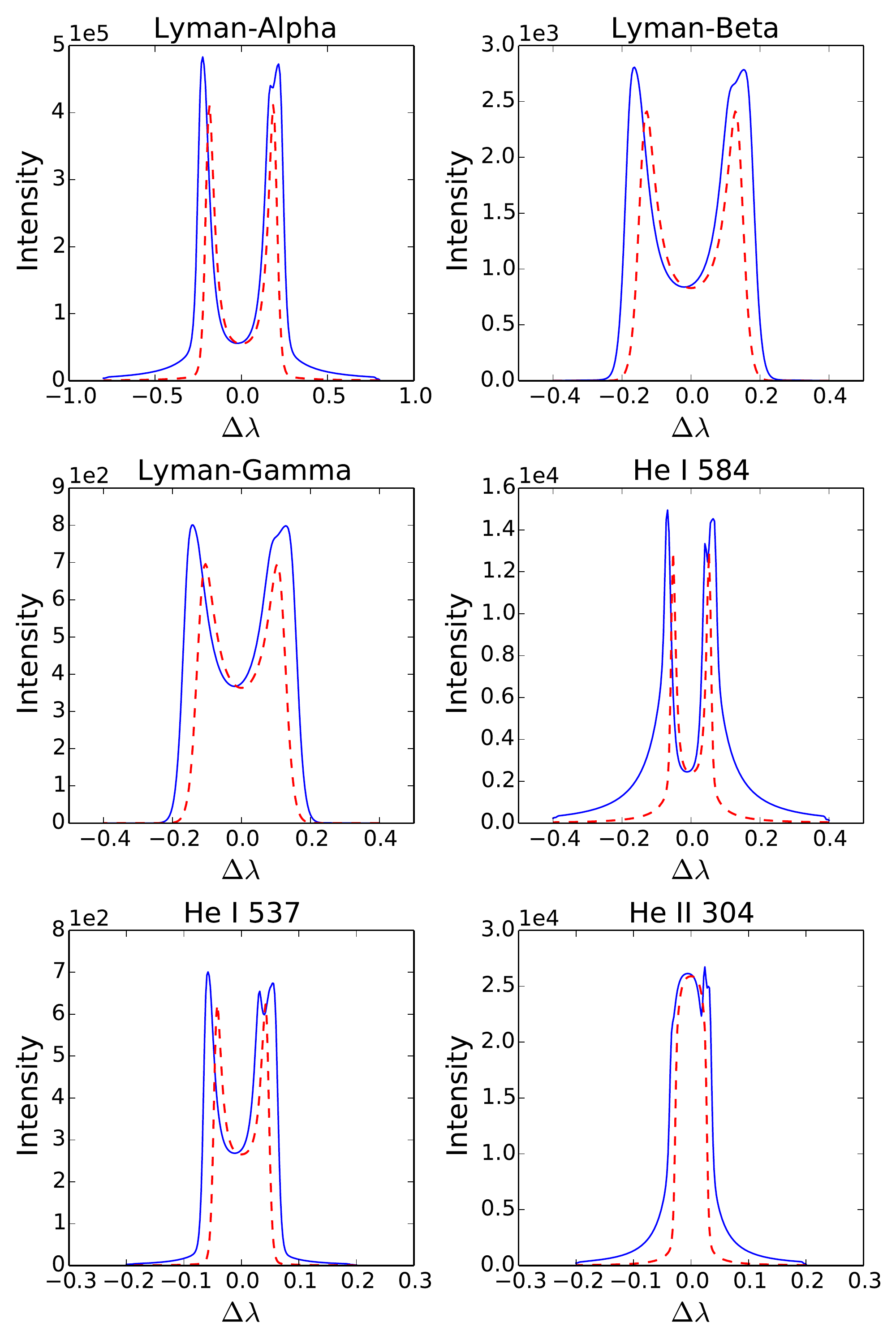}
                \caption{Asymmetrical line profiles of various optically thick hydrogen and helium spectral lines from four unaligned threads with randomly generated velocities. The blue line shows the  profile {after integration through all four moving threads}. The dashed red line shows the profile of a singular, stationary thread. The intensity is given in erg $\mathrm{s^{-1}~cm^{-2}~sr^{-1}~\AA^{-1}}$. {The number at the top left-hand side of each panel is a multiplication factor for the intensity values.} The horizontal axis gives the distance from line centre $\Delta\lambda$ in \AA. The intensities are spatially averaged over the full field of view defined in Fig.~\ref{fig:vlos_diagram}.}
                \label{fig:asymmetries}
        \end{center}
\end{figure}

Each thread in the multi-thread model was assigned a randomly produced LOS velocity in a $\pm 10$~km s$^{-1}$ interval. These velocities are, for the purposes of this study, ad hoc and no physical reason for them is given as of
yet. Assuming that the only change caused by the LOS velocities is the Doppler shift attributed to it, the stationary values for the emergent intensity and optical thickness change according to the thread's Doppler shift, 

\begin{equation}\label{eq:Ichange} 
        I_i(\lambda) \to I_i(\lambda - \Delta \lambda_i) \ ,
\end{equation}
and
\begin{equation}\label{eq:tauchange}
        \tau_i(\lambda) \to \tau_i(\lambda - \Delta \lambda_i) \ .
\end{equation}
The  randomised LOS velocities used in Section~\ref{sec:vlos_results} were specifically $\{u_i\}$=\{7, -7, 9, -2, -7, -9, 4, -3, 4, -7\} km $\mathrm{s^{-1}}$. Each thread was displaced by a random number of position array points and the displacements used were \{0, -119, -40, 120, 9, -45, 55, -86, 37, 3\}, where the total number of array points was 201 covering a thread of diameter specified by the model  {used}, e.g. a \emph{p} model would have displacement values of \{0, -1190, -400, 1200, 900, -450, 550, -860, 370, 30\} km.  {For models with less than ten threads the same velocity and displacement arrays were used with the values past the number of threads ignored.} A schematic diagram of the multi-thread system with velocities in the LOS direction can be seen in Fig.~{\ref{fig:vlos_diagram}. The model used throughout this section is the \emph{p}4 model (see Table~\ref{table:parameters}).

\subsection{{Line profiles} from multi-thread model with LOS velocities}\label{sec:vlos_results}

\begin{figure}
        \begin{center}
                \includegraphics[width=\columnwidth]{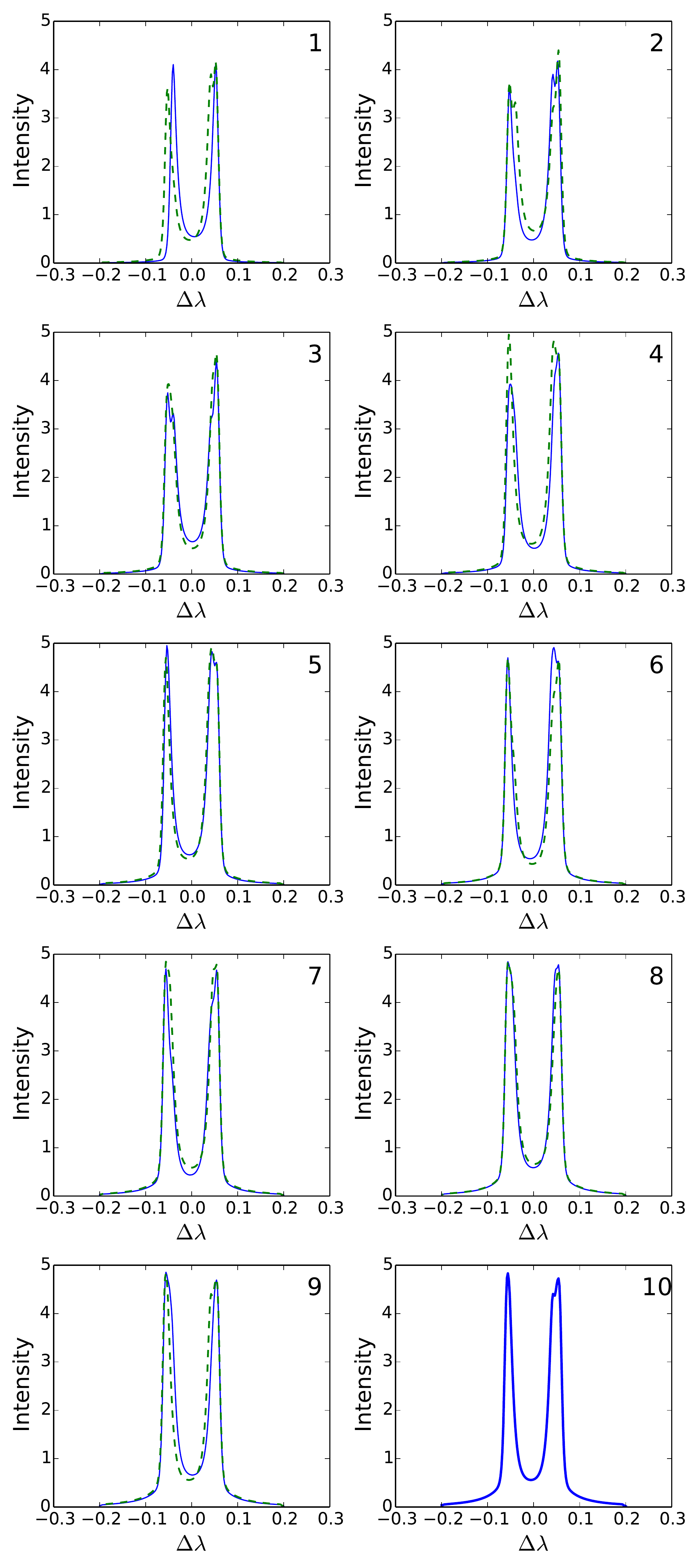}
                \caption{Cumulative effect of {randomly moving, randomly positioned threads} on the {position-averaged} emergent intensity for the Lyman $\alpha$ line in a multi-thread model. The solid blue line shows the emergent intensity after $N$ threads. The number of threads is indicated at the top right of each panel. The dashed green line shows the line profile of the $(N+1)^{\mathrm{th}}$ thread in the LOS. The intensity is given in $10^5$~erg $\mathrm{s}^{-1}~\mathrm{cm}^{-2}~\mathrm{sr}^{-1}~\AA^{-1}$. The distance from line centre $\Delta\lambda$ is given in \AA. {The radiation emergent from each thread is computed with the \emph{p}4 model.}}
                \label{fig:cumulative_asymmetry}
        \end{center}
\end{figure}

To obtain the results seen in this section, the \emph{p}4 model was {used in} a system of unaligned threads, each thread having a LOS velocity as quoted in the previous section. {The number of threads in this system was varied between two and ten.} The results for a selection of hydrogen and helium lines {using a} four thread {model} are given in Fig.~\ref{fig:asymmetries}. 
{This figure shows that }after four threads, the optically thick lines show significant asymmetries.
Fairly complex line profiles result from the superposition of several threads along the LOS.

The cumulative effect {of integrating through one to ten} moving threads on the line profile of the Lyman~$\alpha$ line is shown in Fig.~\ref{fig:cumulative_asymmetry}, which  shows the  {resulting} line profile {averaged over} all positions across the cylinder. By averaging  across the cylinder, much of the complexity of the profiles is smoothed over.  Asymmetries and complex profiles are a feature seen in most profiles after a small number of threads in the LOS direction. However as the number of threads increases, the line profiles tend towards a more symmetric position-averaged profile. 

Some complexity remains, however, in the position-dependent profiles. Figure 12 shows graphs showing position-dependent line profiles for the unaligned, multi-thread model with LOS velocities  for the \ion{He}{ii} 304~\AA\ line; see also Figs.~\ref{fig:lyA_asymmetry}--\ref{fig:heI537_asymmetry}. These figures show, for a variety of hydrogen and helium transition lines, significant variations in the line profiles and in the asymmetries with vertical position across the field of view. For example, the blue-to-red peak ratio changes with vertical position. This is particularly notable in \ion{He}{ii} 304~\AA\ (Fig.~\ref{fig:heII304_asymmetry}).}

\begin{figure*}
\centering
\includegraphics[width=\textwidth]{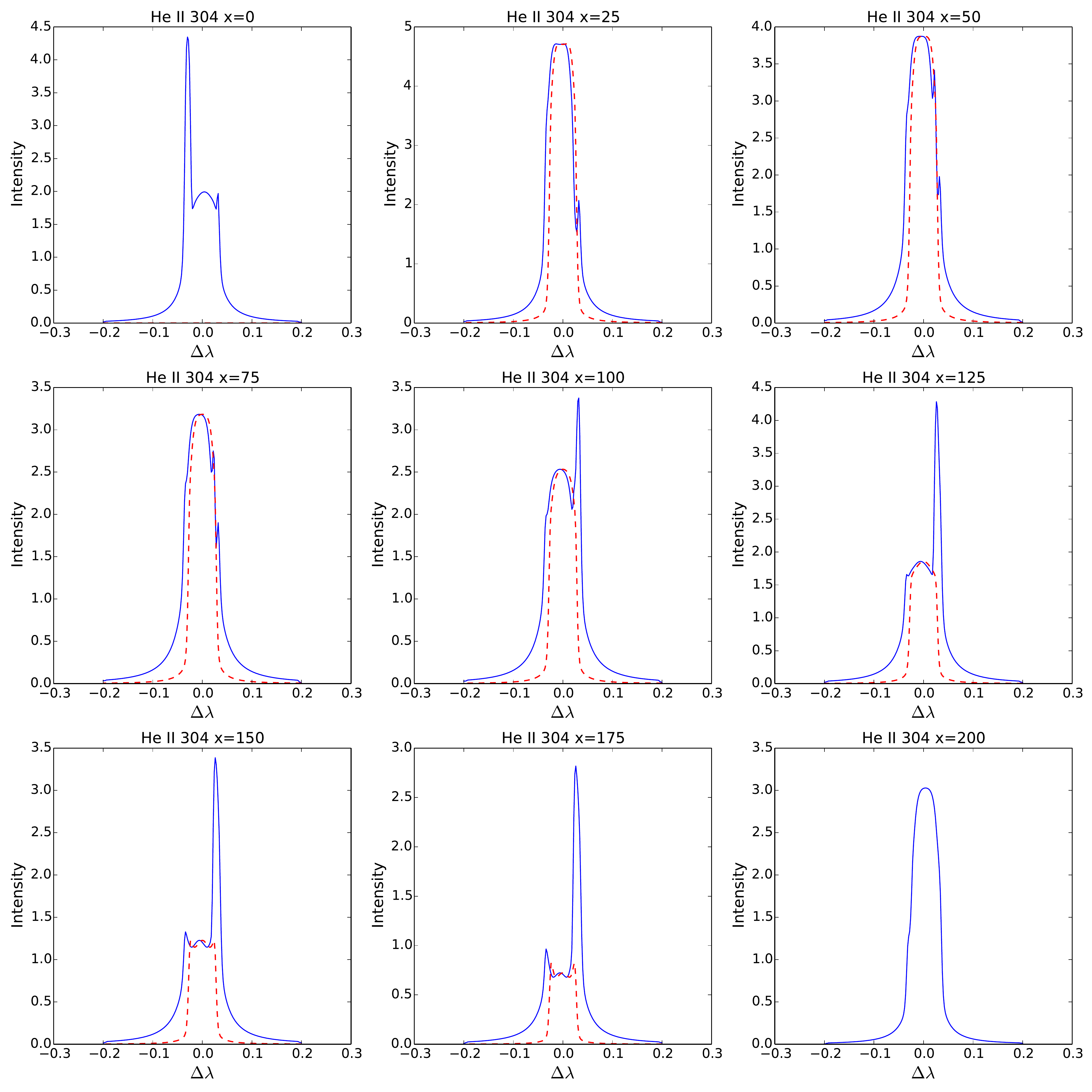}
\caption{Asymmetrical line profiles of \ion{He}{ii} 304~\AA ~line at nine positions across the 201-point field of view. The blue line shows the cumulative profile from ten unaligned threads with randomly generated velocities. The dashed red line shows the profile of a singular, stationary thread. The intensity is given in $10^4$ erg $\mathrm{s^{-1}~cm^{-2}~sr^{-1}~\AA^{-1}}$. The horizontal axis gives the distance from line centre $\Delta\lambda$ in \AA.}
\label{fig:heII304_asymmetry}
\end{figure*}

{The line profiles we show have not been convolved with any  instrumental profile. Such a convolution is necessary to directly compare these models with any observational data. This will result in profiles that are smoothed, attenuating some of the features seen here.}

\section{Discussion and conclusions}\label{sec:conclusions}

The method presented can be successfully used to model a system of multiple {horizontal} solar {chromospheric or} coronal structures {representative of cool loops or prominence threads visible in hydrogen and helium lines}. {The 2D, multi-thread fine structure modelling that combines the hydrogen and helium radiative transfer allows us to compute synthetic emergent spectra from cylindrical structures, and, for the first
time, to study the effect of line-of-sight integration of an ensemble of threads under a range of physical conditions }.

{We show that the presence of a temperature gradient reduces the relative importance of the incident radiation coming from the solar disc on the emergent intensities of most hydrogen and helium lines. The anisotropic irradiation from the solar disc remains important for lines that are primarily formed by resonant scattering of the incident photons, such as for the \ion{He}{ii} 304~\AA\ line.}

When assuming randomly displaced threads in a given field of view, the {variations of the} intensities of optically thick and thin {lines along the vertical direction} are considerably different. In optically thin lines, the emergent intensity increases proportionally with the number of {threads} and the spatial {variation of the} intensity becomes increasingly homogeneous. {The total integrated intensity in} optically thick lines  saturates after only a few {threads, however}. {In that case, the main features seen in the spatial {variation of the} intensity are due to the physical parameters in the foremost threads}. It is also found that  optically thick lines saturate more quickly with increasing {gas} pressure. This is  {related to a general trend of decrease in line width with increasing pressure in optically thick lines. The decreased line width implies that most of the photons in the line are emitted at frequencies where the plasma is optically thick, leading to more absorption and hence greater saturation of the integrated intensity in the LOS.} 

The aim of the analysis performed on these multi-thread models is to explore the possibility of new spectral diagnostics. A result that potentially could be  used is the differences seen between optically thick and optically thin lines for the unaligned, multi-thread model. The emergent intensities from optically thin lines increase with the number of threads, whilst the intensities from optically thick lines become saturated. By comparing the observed and expected intensities from optically thin and thick lines, it may be possible to estimate the number of threads along the LOS {if the plasma parameters are otherwise known}.

This multi-thread model is used to create asymmetric and more complex line profiles through the addition of peculiar velocities {for each thread} along the LOS direction. With randomly generated velocities, the line profiles exhibit significant asymmetries after integration along a small number of threads. With larger numbers of threads, the position-averaged profiles tend towards symmetry, while the position-dependent profiles still exhibit strong asymmetries. 

The same approach may be used to model more realistic {structures}. {For example}, the cylinders radii and orientation with respect to each other, along with their internal plasma parameters, {may} be arbitrarily varied. 
{It may also be interesting to have} a global temperature distribution across the entire multi-thread system, {for example, with cool, isothermal threads surrounded by hotter isothermal threads} {\citep[see e.g.][who compared SUMER observations in Ly-$\alpha$ with multi-thread models using non-identical threads with LOS velocities and with variable orientation with respect to the  LOS]{2015A&A...577A..92S}}. 
{We  plan to carry out a  study {similar to this} using thinner threads representing fine structures of a characteristic size lower than the spatial resolution of current observations.}
This could be used to model bundles of {unresolved} loops.   

{\begin{acknowledgements}
  The authors are grateful to the anonymous referee for constructive comments, which improved the clarity of this paper.
  NL acknowledges support from STFC grant  ST/L000741/1.
  AR acknowledges support from a STFC studentship.
  This research has made use of NASA's Astrophysics Data System.
\end{acknowledgements}
}

%
%



\Online
\begin{appendix}
\section{Position-dependent  line profiles for multi-thread models with LOS velocities}

\begin{figure*}[!H]
\centering
\includegraphics[width=\textwidth]{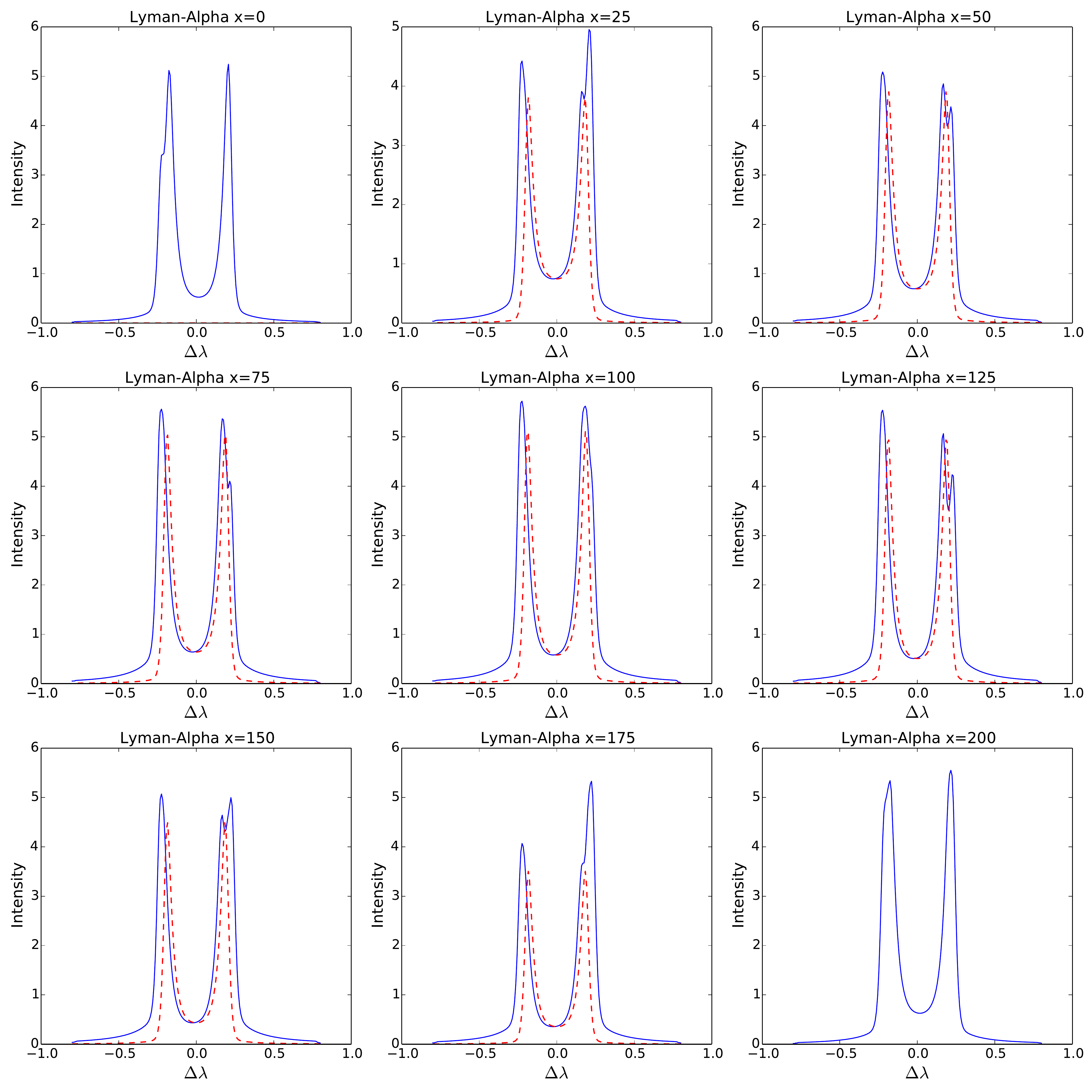}\caption{Asymmetrical line profiles of hydrogen Lyman alpha line at nine positions across the 201-point field of view. The blue line shows the cumulative profile from ten unaligned threads with randomly generated velocities. The dashed red line shows the profile of a singular, stationary thread. The intensity is given in $10^5$ erg $\mathrm{s^{-1}~cm^{-2}~sr^{-1}~\AA^{-1}}$. The horizontal axis gives the distance from line centre $\Delta\lambda$ in \AA.}
\label{fig:lyA_asymmetry}
\end{figure*}

\begin{figure*}
\centering
\includegraphics[width=\textwidth]{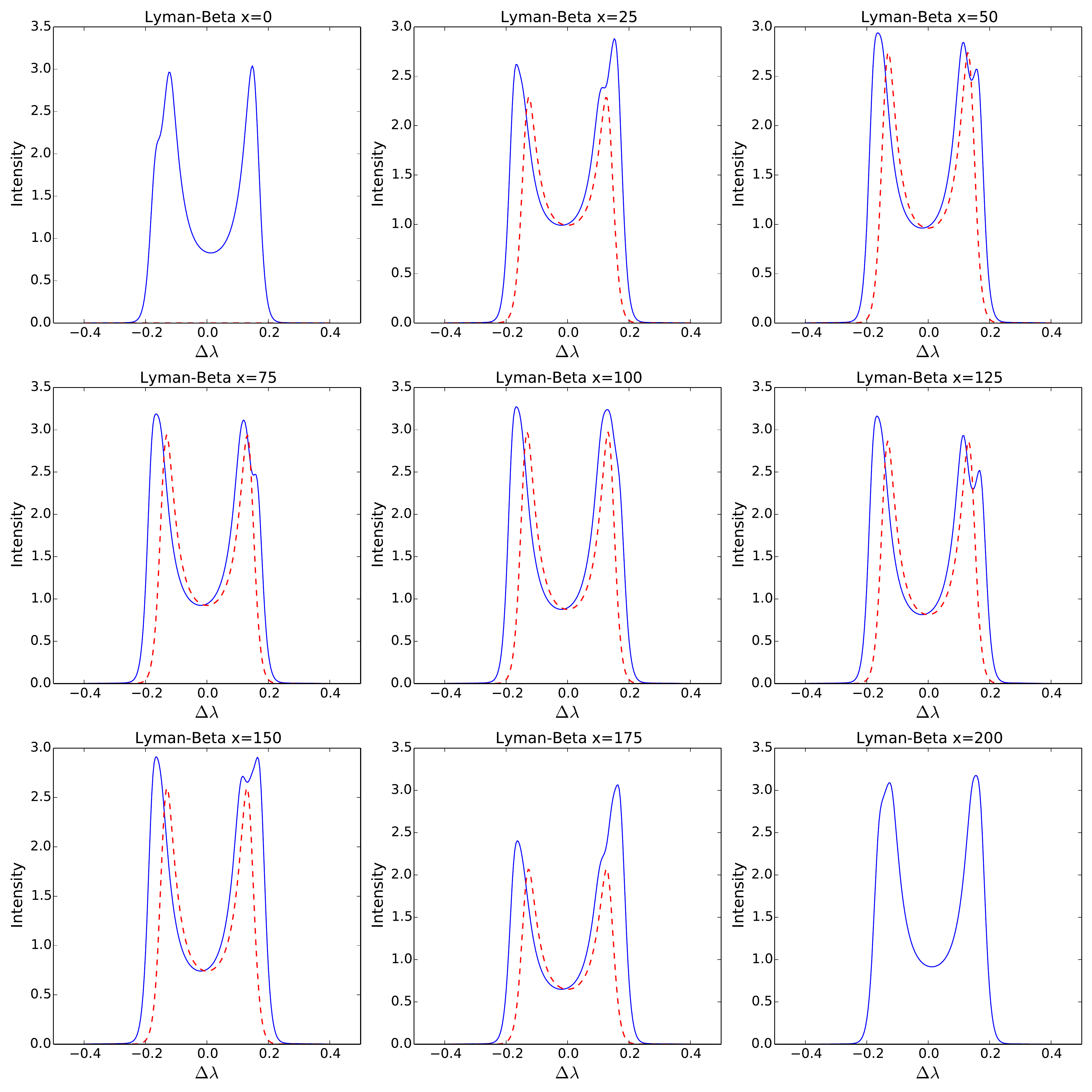}
\caption{Asymmetrical line profiles of hydrogen Lyman beta line at nine positions across the 201-point field of view. The blue line shows the cumulative profile from ten unaligned threads with randomly generated velocities. The dashed red line shows the profile of a singular, stationary thread. The intensity is given in $10^3$ erg $\mathrm{s^{-1}~cm^{-2}~sr^{-1}~\AA^{-1}}$. The horizontal axis gives the distance from line centre $\Delta\lambda$ in \AA.}
\label{fig:lyB_asymmetry}
\end{figure*}

\begin{figure*}
\centering
\includegraphics[width=\textwidth]{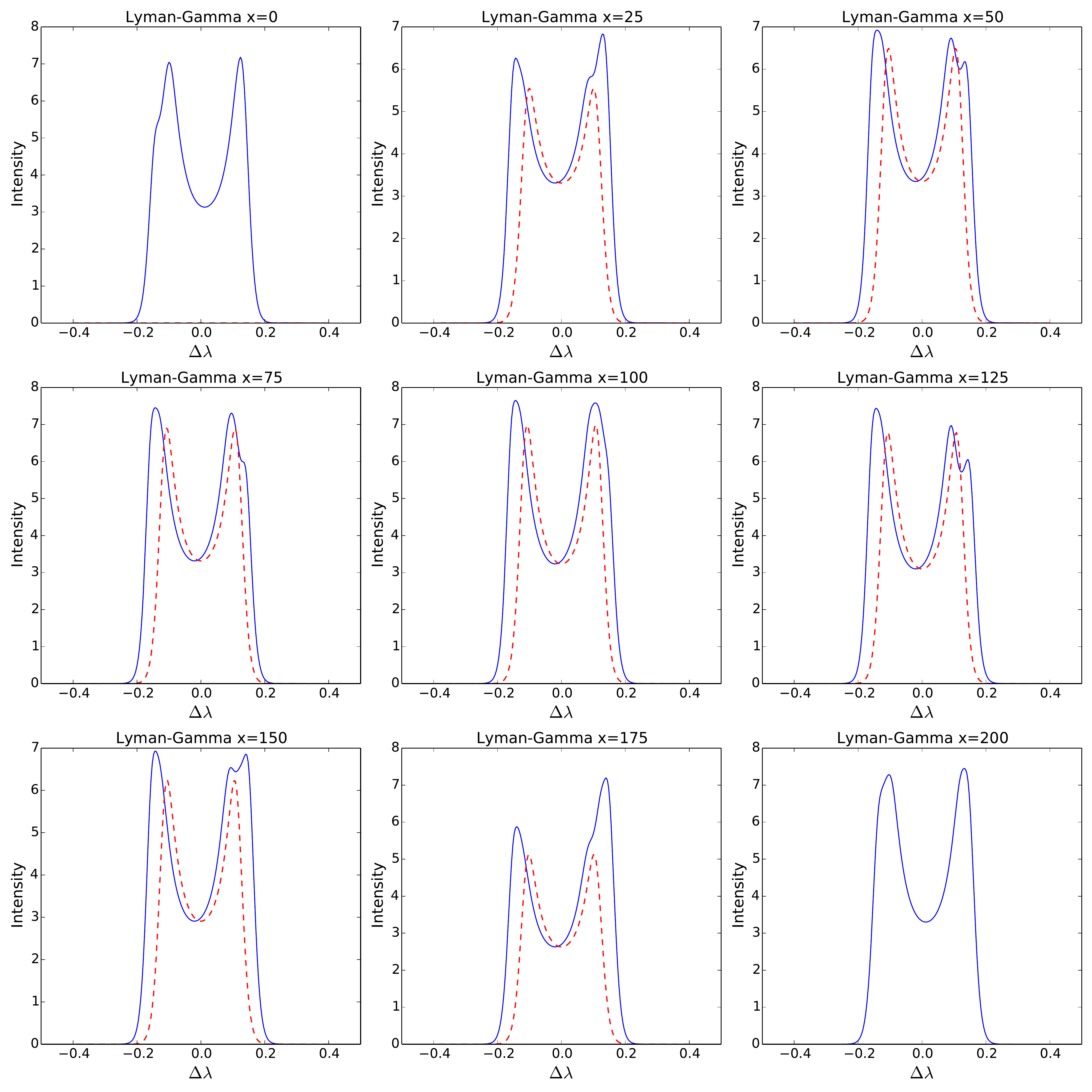}\caption{Asymmetrical line profiles of hydrogen Lyman gamma line at nine positions across the 201-point field of view. The blue line shows the cumulative profile from ten unaligned threads with randomly generated velocities. The dashed red line shows the profile of a singular, stationary thread. The intensity is given in $10^2$ erg $\mathrm{s^{-1}~cm^{-2}~sr^{-1}~\AA^{-1}}$. The horizontal axis gives the distance from line centre $\Delta\lambda$ in \AA.}
\label{lyG_asymmetry}
\end{figure*}

\begin{figure*}
\centering
\includegraphics[width=\textwidth]{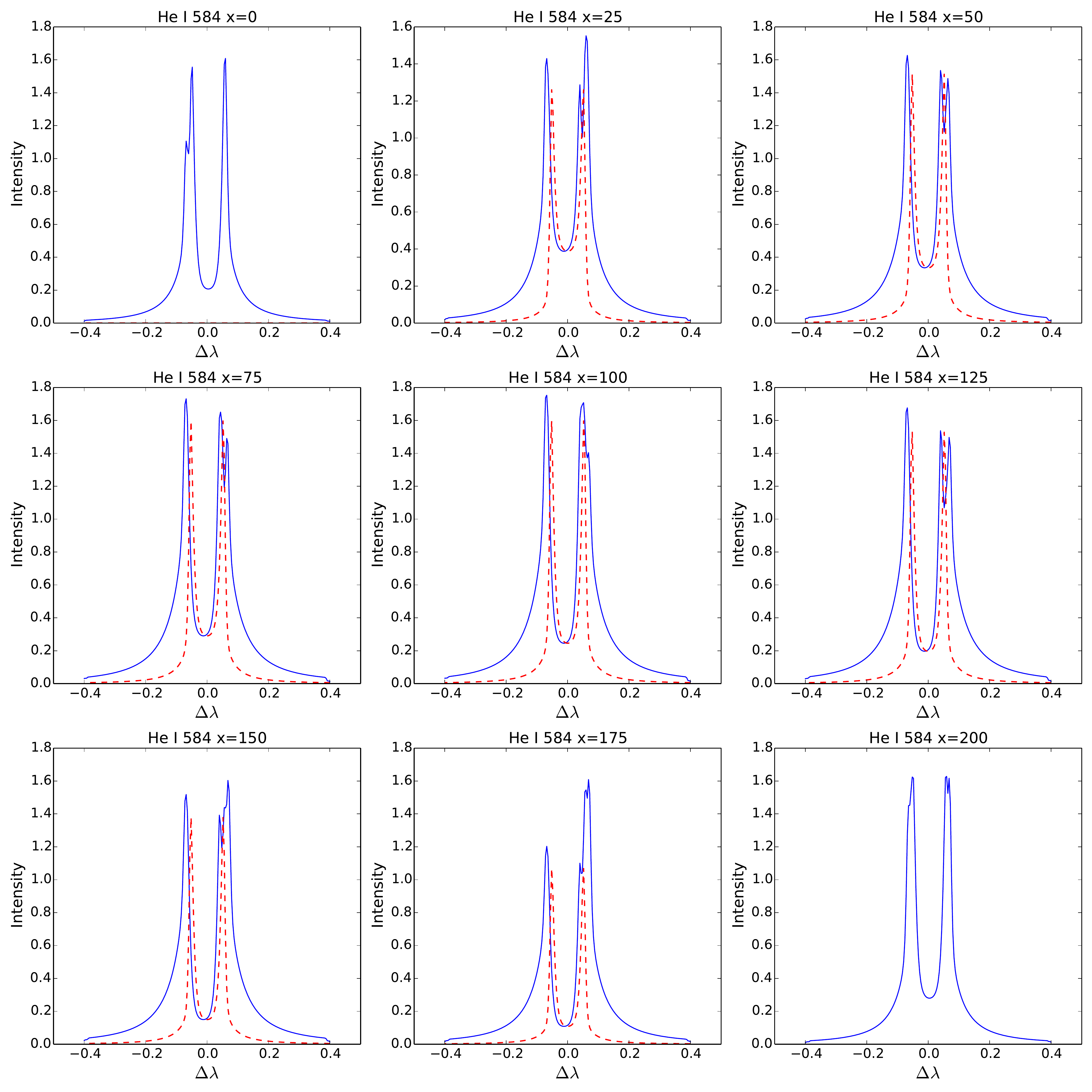}
\caption{Asymmetrical line profiles of \ion{He}{i} 584~\AA ~line at nine positions across the 201-point field of view. The blue line shows the cumulative profile from ten unaligned threads with randomly generated velocities. The dashed red line shows the profile of a singular, stationary thread. The intensity is given in $10^4$ erg $\mathrm{s^{-1}~cm^{-2}~sr^{-1}~\AA^{-1}}$. The horizontal axis gives the distance from line centre $\Delta\lambda$ in \AA.}
\label{fig:heI584_asymmetry}
\end{figure*}

\begin{figure*}
\centering
\includegraphics[width=\textwidth]{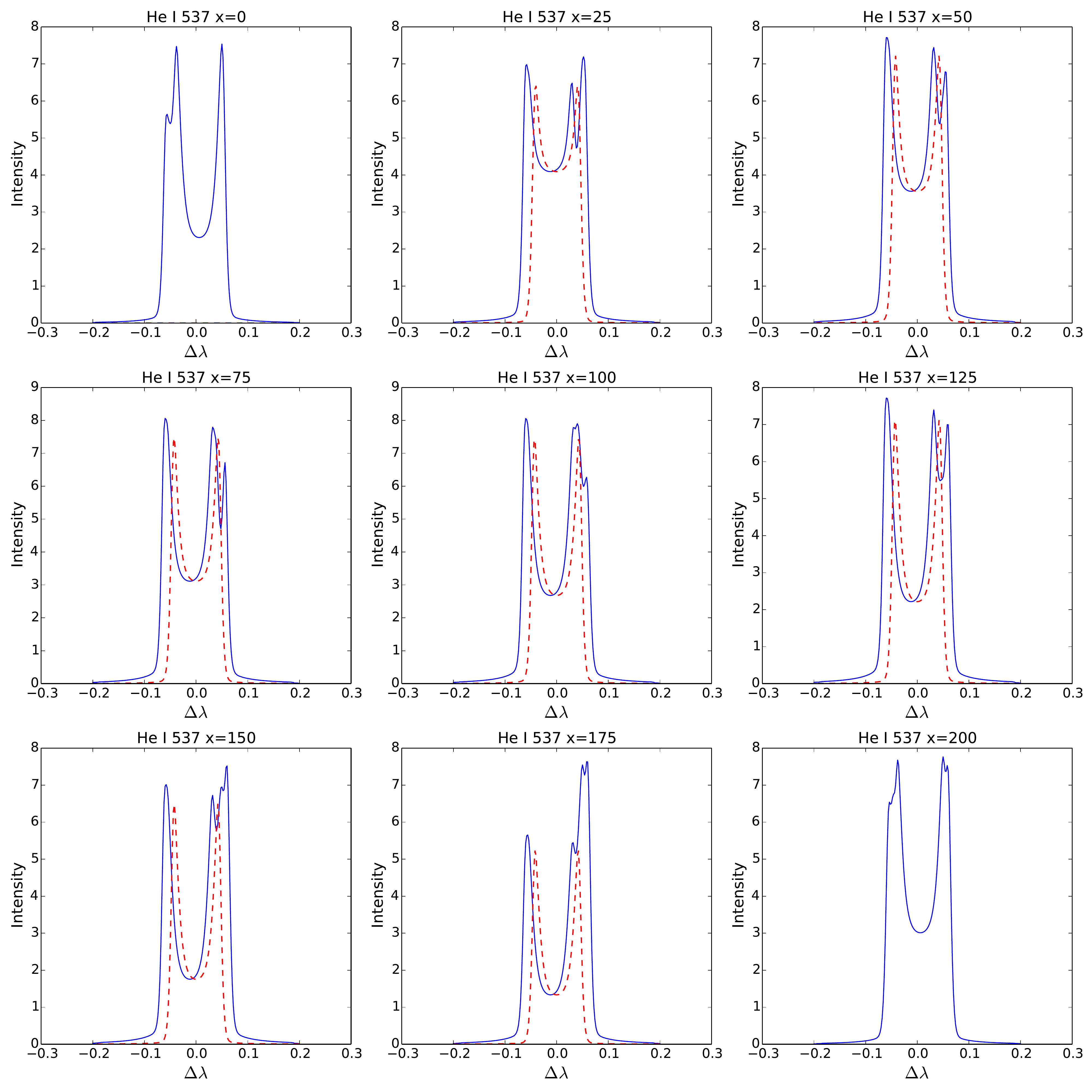}
\caption{Asymmetrical line profiles of \ion{He}{i} 537~\AA ~line at nine positions across the 201-point field of view. The blue line shows the cumulative profile from ten unaligned threads with randomly generated velocities. The dashed red line shows the profile of a singular, stationary thread. The intensity is given in $10^2$ erg $\mathrm{s^{-1}~cm^{-2}~sr^{-1}~\AA^{-1}}$. The horizontal axis gives the distance from line centre $\Delta\lambda$ in \AA.}
\label{fig:heI537_asymmetry}
\end{figure*}

\end{appendix}

\end{document}